\documentclass[preprint,12pt]{elsarticle}

\usepackage{amssymb}
\usepackage{amsmath}
\usepackage{hyperref}
\usepackage{comment}
\usepackage{subcaption}
\usepackage{ushort}
\usepackage{mathabx}
\usepackage{tikzpeople}
\usepackage{pgfplots}
\usepackage{xurl}

\usepackage{booktabs}
\usepackage{tabularx}
\usepackage{array}
\usepackage{float}
\usepackage{makecell}

\pgfplotsset{compat=1.18}

\newcounter{defcounter}
\allowdisplaybreaks

\journal{Electric Power Systems Research}

\begin{document}

\begin{frontmatter}



\title{A Tax-Subsidy Scheme for Efficient Investment in Renewable Generation Capacity}


\author{Mohammad Reza Karimi Gharigh}
\author{Lamia Varawala}
\author{Mohammad Reza Hesamzadeh}
\author{Gy\"orgy D\'an}

\affiliation{organization={Department of Electrical Engineering and Computer Science, KTH Royal Institute of Technology},
            addressline={Teknikringen 33}, 
            postcode={114 28}, 
            city={Stockholm},
            country={Sweden}}

\begin{abstract}
Electricity markets are central to the energy transition, but current market rules do not always deliver socially desirable outcomes. This paper studies two related inefficiencies in wholesale electricity markets. First, competitive market-clearing based on merit-order dispatch can ignore pollution damages, leading to inefficient spot-market outcomes. We address this distortion by deriving a Pigouvian tax that charges each generator for its marginal contribution to pollution damages. Second, because generation capacity affects future spot prices, producers may strategically distort their long-run investment decisions, particularly in systems with increasingly low-operating-cost renewable generation. To align private investment incentives with social welfare, we derive a generator-specific investment subsidy based on each producer's contribution to aggregate utility. We show that, under the stated behavioral and informational assumptions, the combined tax--subsidy mechanism aligns decentralized dispatch and generation-capacity investment decisions with the social-welfare benchmark while respecting operational constraints. Implementing this mechanism does not require producers to report their complete private cost functions; the tax and subsidy can instead be calculated using existing market submissions and settlement data, verified emission information, and generator-specific counterfactual market re-clearing outcome. An analytical example and numerical experiments illustrate the mechanism and show that strategic capacity investment can exceed the socially optimal level and reduce welfare.
\end{abstract}


\end{frontmatter}



\section{Introduction}
\label{sec:introduction}
Wholesale electricity markets (WEMs) clear daily to balance supply and demand. They are typically managed by an independent system operator (ISO), which performs competitive market clearing and merit-order dispatch subject to system constraints~\citep{biggar2014economics}. This design works well for minimizing production costs, but it does not account for the resulting pollution, such as carbon emissions. This misalignment conflicts with the climate targets of many countries~\citep{un}, such as Great Britain, where the Office of Gas and Electricity Markets (Ofgem) is required to support the UK’s 2050 net-zero target~\citep{ofgem_netzero_2023}, and the EU, where recent reforms aim to reduce price risk and support long-term investment in renewable generation~\citep{eu_em_reform_2024}.

Addressing these environmental objectives within markets originally designed for short-run efficiency is intrinsically challenging, owing to a fundamental tension between short-run efficiency and long-run investment incentives. In standard spot-market clearing, prices reflect individual operation costs, while pollution damages are not accounted for. Because pollution remains an externality, optimal dispatch can be socially inefficient, and emissions can be high. Carbon policies attempt to address this through cap-and-trade systems or carbon taxes~\citep{hua2024carbon, akulker2023optimal, carbon_tax}, and carbon futures markets are widely used for hedging and price discovery, especially in the EU~\citep{xi2023re}. However, carbon policies also interact with market rules such as price controls and cross-border trade, which can affect both prices and renewable investment incentives~\citep{carlos2024effect}. In addition, strategic behavior in operations can amplify inefficiencies. For instance, firms may withhold generation capacity or exploit technical limits to raise prices~\citep{hesamzadeh2012merger, moiseeva2017generation}, and it can be difficult to distinguish strategic withholding from true outages~\citep{RePEc:hhs:iuiwop:1015}. Regulators can respond with competition rules and safeguards~\citep{strategic_laws}, and price caps have also been proposed as a remedy~\citep{price_cap_propose}.

The second challenge concerns long-run investment. Renewable generation is expanding rapidly, and its operation costs have fallen in many places~\citep{irena_cost_2023, iea_renewables_2024}. Because renewable technologies have low operation costs, increasing renewable capacity can reduce wholesale spot prices and producer profits. This can weaken incentives to invest in clean generation, such as solar and wind, especially when investment costs are high. For this reason, many systems use long-term contracts such as contracts for differences (CfDs) and power purchase agreements (PPAs)~\citep{bruninx2022covid, mesa2023long}, as well as generation capacity mechanisms~\citep{kozlova2023interface, holmberg2023survey}. Still, investment is difficult to control directly, and strategic generation expansion has been studied using equilibrium and bi-level models in which firms anticipate how generation capacity affects dispatch and prices~\citep{tohidi2016sequential}. Recent work also argues that externalities and market power should be addressed jointly. For instance, \citep{incentive} proposes a pricing mechanism that targets both market power and environmental externalities and considers the ISO’s information limits. A related debate is whether current market designs provide sufficient long-run incentives, and whether generation capacity markets address the underlying problem~\citep{biggar_hesamzadeh_capacity_2025}. References~\citep{biggar_hesamzadeh_capacity_2025} and \citep{hesamzadeh_biggar} link this debate to risk allocation, hedging, and long-term contracting.

Existing studies address either the short-run electricity-market dispatch with the pollution externality or the long-run generation-capacity investment, but generally do not address both within a single corrective mechanism. Carbon-pricing studies focus primarily on internalizing pollution damages in market operations. A Pigouvian tax can therefore correct dispatch incentives, but it does not by itself correct the effect of generation capacity on producers' long-run investment incentives. Studies of strategic generation expansion examine how capacity affects dispatch and market prices, but do not jointly derive a pollution tax and an investment-support instrument that align both decisions with a common social-welfare benchmark. Similarly, mechanisms addressing operational market power and environmental externalities generally do not incorporate strategic generation-capacity investment. The research gap is therefore the design of a mechanism that simultaneously corrects the pollution externality in short-run dispatch and the distortion of capacity investment on market prices in long-run investment. Table~\ref{tab:literature_comparison} summarizes this gap.

\begin{table}[t]
    \centering
    \caption{Comparison of the present study with related literature.}
    \label{tab:literature_comparison}

    \scriptsize
    \setlength{\tabcolsep}{2.5pt}
    \renewcommand{\arraystretch}{1.20}

    \begin{tabularx}{\textwidth}{
        >{\raggedright\arraybackslash}p{1.55cm}
        >{\centering\arraybackslash}p{1.45cm}
        >{\centering\arraybackslash}p{1.45cm}
        >{\centering\arraybackslash}p{1.30cm}
        >{\centering\arraybackslash}p{1.55cm}
        >{\centering\arraybackslash}p{1.55cm}
        >{\centering\arraybackslash}X
    }
        \toprule
        \textbf{Reference}
        & \makecell{\textbf{Pollution}\\\textbf{policy}}
        & \makecell{\textbf{Market}\\\textbf{operation}}
        & \makecell{\textbf{Strategic}\\\textbf{behavior}}
        & \makecell{\textbf{Capacity}\\\textbf{investment}}
        & \makecell{\textbf{Capacity-induced}\\\textbf{price effect}}
        & \makecell{\textbf{Corrective investment}\\\textbf{support}} \\
        \midrule
        \cite{HESAMZADEH20149}                 & No        & Yes       & No        & No        & No        & No \\
        \cite{hesamzadeh_biggar}               & No        & Yes       & No        & No        & No        & No \\
        \cite{price_cap_propose}                & No        & No        & Partially & No        & No        & No \\
        \cite{hesamzadeh2012merger}             & No        & Yes       & Yes       & No        & No        & No \\
        \cite{RePEc:hhs:iuiwop:1015}            & No        & Yes       & Yes       & No        & No        & No \\
        \cite{moiseeva2017generation}           & No        & Yes       & Yes       & No        & No        & No \\
        \cite{incentive}                        & Yes       & Yes       & Yes       & No        & No        & No \\
        \cite{tohidi2016sequential}             & No        & Yes       & Yes       & Yes       & Partially & No \\
        \cite{biggar_hesamzadeh_capacity_2025}  & No        & Partially & Partially & Yes       & No        & Yes \\
        \cite{holmberg2023survey}               & Partially & Partially & Partially & Yes       & No        & Yes \\
        \cite{kozlova2023interface}             & Partially & No        & No        & Partially & No        & Yes \\
        \cite{mesa2023long}                     & No        & No        & No        & No        & No        & Partially \\
        \cite{bruninx2022covid}                 & Yes       & No        & No        & No        & No        & No \\
        \cite{xi2023re}                         & Yes       & No        & No        & No        & No        & No \\
        \cite{carbon_tax}                       & Yes       & Partially & Partially & Partially & No        & No \\
        \cite{akulker2023optimal}               & Yes       & Partially & No        & Partially & No        & No \\
        \cite{carlos2024effect}                 & Yes       & Yes       & Yes       & Yes       & Partially & No \\
        \cite{hua2024carbon}                    & Yes       & No        & Partially & Partially & No        & No \\
        \midrule
        \textbf{Present study}                  & \textbf{Yes} & \textbf{Yes} & \textbf{Yes} & \textbf{Yes} & \textbf{Yes} & \textbf{Yes} \\
        \bottomrule
    \end{tabularx}
\end{table}

The main novelty is not simply the simultaneous use of a pollution tax and an investment subsidy. Both instruments are derived analytically from a common comparison of decentralized and social-welfare optimality conditions. The Pigouvian tax corrects the marginal pollution externality in short-run dispatch, whereas the generator-specific investment support corrects the effect of capacity investment on market prices in strategic long-run investment. The framework, therefore, identifies and corrects the distinct distortions affecting dispatch and generation-capacity investment.

The proposed mechanism addresses only one aspect of the transition toward a reliable and decarbonized power system. Its effectiveness also depends on complementary developments in transmission networks, system flexibility, operational reliability, and smart grid infrastructure \citep{adeleke2025comprehensive}. Advanced metering and internet of thing-based monitoring can support implementation by providing system data over time. The real-time advanced metering infrastructure developed by \citep{kim2025design} illustrates this enabling technological layer, although additional market, network, emissions, and counterfactual information remains necessary.

Based on the identified research gap, this paper makes three main contributions. First, we derive a generator-level Pigouvian tax based on each generator's marginal contribution to pollution damages. Under the stated assumptions, the tax aligns decentralized dispatch with the socially optimal dispatch. Second, we identify the effect of capacity investment on market prices, which causes private investment incentives to diverge from the socially optimal investment incentives. We then derive a generator-specific investment subsidy that corrects this term and aligns the producer's capacity decision with the social-welfare benchmark. Third, the mechanism does not require new producer-specific reports of complete private cost functions beyond the information already used in market clearing and settlement. It nevertheless requires verified emissions and technical data, regulator-specified pollution-damage functions, and generator-specific counterfactual market re-clearing.\footnote{The information structure is inspired by the incentive approach in \citep{hesamzadeh2018simple} and related work such as \citep{khastieva2020transmission}.} Unlike studies that address pollution pricing, operational market power, or capacity investment separately, the proposed framework coordinates short-run dispatch and long-run generation-capacity investment within the same market model.

The remainder of the paper is organized as follows. Section~\ref{sec:spot_market} presents the WEM model and derives the Pigouvian tax. Section~\ref{sec:investment} studies generation-capacity investment and derives the investment subsidy. Section~\ref{sec:properties} discusses the properties and implementation requirements of the proposed tax--subsidy mechanism. Sections~\ref{sec:example} and~\ref{sec:experiments} present an analytical example and numerical experiments, respectively. Finally, Section~\ref{sec:conclusion} concludes the paper.
\vspace{-0.3cm}
\section{Spot-market generation and Pigouvian taxation}
\label{sec:spot_market}

This section examines spot-market dispatch and derives the corrective Pigouvian tax. We first formulate the social planner's dispatch problem, which includes pollution damages in social welfare. We then formulate a competitive market-clearing price, in which these damages are omitted. Comparing the corresponding optimality conditions identifies the missing marginal-damage term and yields a generator-specific tax that aligns competitive dispatch with the social-welfare benchmark.

The analysis adopts a two-stage behavioral benchmark. Conditional on installed generation capacities, the spot market is competitively cleared by the independent system operator (ISO), and producers are price-takers in the dispatch stage. Operating-cost bids and relevant technical limits are assumed to be truthful or independently verifiable. The ISO has access to the demand, network, bid, and technical information ordinarily used in market clearing. In the investment stage, however, producers form a finite set of generation firms whose capacity decisions can affect residual supply, network congestion, and future equilibrium prices. They therefore account for the effect of capacity investment on market prices when making long-run investment decisions. Competitive dispatch conditional on installed capacity does not imply price-taking behavior in generation-capacity investment.

The separation of long-run investment decisions from short-run market operation is consistent with established generation-investment models \citep{hesamzadeh2011generation, wogrin2012capacity, kallabis2023strategic}. In particular, two-stage capacity-expansion models distinguish investment decisions from subsequent market outcomes, and some formulations allow short-run behavior to range from perfect competition to more strategic market conduct \citep{wogrin2012capacity}. This separation also has an institutional basis: in liberalized electricity systems, generation investment is decentralized among private firms, whereas short-run market clearing is conducted by a market operator or ISO \citep{kallabis2023strategic}. Here, competitive dispatch is used as a post-mitigation benchmark to isolate the effect of capacity investment on market prices in the long-run investment decisions. Allowing strategic behavior in both dispatch and investment would introduce additional distortions and a substantially more complex equilibrium problem, and is left for future research.

The competitive-dispatch assumption is interpreted as a post-mitigation benchmark. This interpretation reflects the institutional separation between decentralized generation investment and centralized short-run market clearing by the ISO and is consistent with generation-investment models that distinguish long-run investment decisions from subsequent spot-market behavior \citep{wogrin2012capacity,kallabis2023strategic}. Standard monitoring and market-power mitigation measures, such as conduct-and-impact tests \citep{de2004use}, are treated as elements of the institutional environment supporting this benchmark rather than as endogenous components of the model. More detailed formulations can explicitly represent dispatch-side market power through strategic price--quantity decisions and equilibrium models \citep{hesamzadeh2009transmission}; however, incorporating such behavior introduces equilibrium-based models and increases the computational complexity of the resulting model. Accordingly, the model does not explicitly represent strategic bidding, economic or physical withholding, or the strategic misreporting of operational constraints. Restricting strategic behavior to the investment stage allows the analysis to isolate the price effects of strategic capacity investment and the associated investment distortion from additional sources of short-run market power. If residual dispatch-side market power remains after monitoring and mitigation, additional distortions arise, and the proposed tax and subsidy alone need not implement the social-welfare benchmark. Jointly modeling market power in short-run operation and long-run investment market power is therefore left for future research.

\vspace{-0.3cm}
\subsection{Socially optimal spot-market generation}
We consider a power system with a set of buses $\mathcal{I}$ connected by a set of transmission lines $\mathcal{L}$. WEM participants include producers indexed by $p\in\mathcal{P}$ and consumers represented by demands indexed by $d\in\mathcal{D}$. Producers own generating units indexed by $g\in\mathcal{G}$, and $\mathcal{G}_p$ denotes the set of units owned by producer $p$. We study dispatch over a planning horizon of $T$ intervals, indexed by $t\in\mathcal{T}=\{1,2,\dots,T\}$.

For generator $g$, the installed generation capacity is $\widebar{P}_{g}^{G}$ (assumed constant over the horizon), and the output in interval $t$ is $P_{g,t}^G$. For renewable technologies (e.g., wind or solar), available generation capacity can change over time. We model this using an availability factor $A_{g,t}^G\in[0,1]$ and a relative output decision $R_{g,t}^G\in[0,1]$. Output is linked to these variables by
\vspace{-0.2cm} 
\begin{equation}
\hspace{-0.4cm}
\label{eq:relative_generation}
    P_{g,t}^G = R_{g,t}^G A_{g,t}^G \widebar{P}_{g}^G \leftrightarrow \omega_{g,t},
    \; \forall g\in\mathcal{G},\; t\in\mathcal{T}.
\end{equation}
\begin{equation}
\hspace{-0.4cm}
\label{eq:generation_capacity_limits}
    0 \leq R_{g,t}^G \leq 1 \leftrightarrow \left(\ushort{\mu}_{g,t},\widebar{\mu}_{g,t}\right),
    \; \forall g\in\mathcal{G},\; t\in\mathcal{T}.
\end{equation}
We also include ramping limits that restrict changes in relative output between consecutive intervals
\vspace{-0.2cm} 
\begin{equation}
\hspace{-0.4cm}
\label{eq:ramping_limits}
    -\widebar{R}_{g}^G \leq R_{g,t}^G - R_{g,t-1}^G \leq \widebar{R}_{g}^G
    \leftrightarrow \left(\ushort{\rho}_{g,t},\widebar{\rho}_{g,t}\right),
    \; \forall g\in\mathcal{G},\; t\in\mathcal{T}\setminus\{1\}.
\end{equation}

Each generator has an operation cost function $\mathbf{C}_g^O(P_{g,t}^G)$, which is non-negative, non-decreasing, and convex in $P_{g,t}^G$. Pollution created at bus $i$ by producer $p$ in interval $t$ is denoted by $X_{p,i,t}$ and depends on the output of units owned by $p$ at bus $i$. We write this as $X_{p,i,t}=\mathbf{X}_{p,i}(P_{g,t}^G)$ for the relevant units, where $\mathbf{X}_{p,i}(\cdot)$ is non-negative, non-decreasing, and convex. Pollution creates damages that depend on total pollution at each bus. We define
\vspace{-0.2cm}
\[
    Z_{i,t} := \sum_{p\in\mathcal{P}} X_{p,i,t}, \; \forall i\in\mathcal{I},\; t\in\mathcal{T},
\]
with the damage function $\mathbf{E}_i(Z_{i,t})$, which is assumed to be non-negative, non-decreasing, and convex in $Z_{i,t}$. On the demand side, demand $d$ consumes $P_{d,t}^{D}$ in interval $t$, bounded above by $\widebar{P}_{d,t}^{D}$
\begin{equation}
\label{eq:demand_limits}
    0 \leq P_{d,t}^{D} \leq \widebar{P}_{d,t}^{D} \leftrightarrow (\ushort{\alpha}_{d,t},\widebar{\alpha}_{d,t}),
    \; \forall d\in\mathcal{D},\; t\in\mathcal{T}.
\end{equation}
Consumers obtain utility $\widetilde{\mathbf{U}}_{d,t}(P_{d,t}^{D})$, which is assumed to be non-decreasing and concave in $P_{d,t}^{D}$. In each interval, total generation must equal total demand
\begin{equation}
\label{eq:power_balance}
    \sum_{d\in\mathcal{D}} P_{d,t}^{D} - \sum_{g\in\mathcal{G}} P_{g,t}^{G} = 0 \leftrightarrow (\lambda_t),
    \; \forall t\in\mathcal{T}.
\end{equation}
We represent transmission constraints using a power transfer distribution factor (PTDF) matrix $H_{l,i}$~\citep{HESAMZADEH20149}. Let $\mathcal{D}_i$ and $\mathcal{G}_i$ denote demands and generators located at bus $i$. Then, line $l$ has flow limits given by
\vspace{-0.2cm}
\begin{equation}
\hspace{-0.4cm}
\label{eq:line_limits}
    -\widebar{F}_l \leq \sum_{i\in\mathcal{I}} H_{l,i}\!\left(\sum_{d\in\mathcal{D}_i} P_{d,t}^{D}-\sum_{g\in\mathcal{G}_i} P_{g,t}^{G}\right)\leq \widebar{F}_l
    \leftrightarrow (\ushort{\beta}_{l,t},\widebar{\beta}_{l,t}),
    \; \forall l\in\mathcal{L},\; t\in\mathcal{T}.
\end{equation}

Given these definitions, the social planner chooses generation and demand to maximize total utility minus operation costs and pollution damages
\vspace{-0.2cm}
\begin{equation}
\hspace{-0.4cm}
\label{eq:Social_welfare_objective}
    \mathbf{W}^M\!\left(\widebar{P}_{g}^{G}\right)
    =
    \max \sum_{t\in\mathcal{T}}\!\left[
        \sum_{d\in\mathcal{D}} \widetilde{\mathbf{U}}_{d,t}(P_{d,t}^{D})
        - \sum_{g\in\mathcal{G}} \mathbf{C}_g^O(P_{g,t}^G)
        - \sum_{i\in\mathcal{I}} \mathbf{E}_i(Z_{i,t})
    \right],
\end{equation}
subject to \eqref{eq:relative_generation}--\eqref{eq:line_limits}.

It is useful to work with an aggregate utility function that already accounts for how the ISO allocates demand across nodes under the network constraints. For each interval $t$, we define the value function
\begin{equation}
\hspace{-0.4cm}
\label{eq:utility}
    \mathbf{U}_t\!\left(P_{g,t}^G\right)
    =
    \max \sum_{d\in\mathcal{D}} \widetilde{\mathbf{U}}_{d,t}(P_{d,t}^{D}),\; \forall t\in\mathcal{T},
\end{equation}
subject to \eqref{eq:demand_limits}--\eqref{eq:line_limits}, where $P_{g,t}^G$ is treated as a fixed value in the above problem. In other words, $\mathbf{U}_t(\cdot)$ returns the maximum total utility that can be achieved for a given pattern of generation, after enforcing demand bounds, power balance, and line limits. Using $\mathbf{U}_t(\cdot)$, we can write the planner’s welfare problem in terms of generation decisions as
\vspace{-0.2cm}
\begin{equation}
\hspace{-0.4cm}
\label{eq:welfare}
    \mathbf{W}^M\!\left(\widebar{P}_{g}^{G}\right)
    =
    \max \sum_{t\in\mathcal{T}}\!\left[
        \mathbf{U}_t(P_{g,t}^G)
        - \sum_{g\in\mathcal{G}} \mathbf{C}_g^O(P_{g,t}^G)
        - \sum_{i\in\mathcal{I}} \mathbf{E}_i(Z_{i,t})
    \right],
\end{equation}
subject to \eqref{eq:relative_generation}--\eqref{eq:ramping_limits}. We denote optimal values by the superscript $(\cdot)^*$. The Karush--Kuhn--Tucker (KKT) stationarity condition with respect to $R_{g,t}^G$ can be written as
\vspace{-0.2cm}
\begin{equation}
\hspace{-0.4cm}
\label{eq:kkt_general}
    A_{g,t}^G \widebar{P}_{g}^G \,\omega_{g,t}^{*}
    =
    - \ushort{\mu}_{g,t}^{*} + \widebar{\mu}_{g,t}^{*}
    - \ushort{\rho}_{g,t}^{*} + \widebar{\rho}_{g,t}^{*}
    + \ushort{\rho}_{g,t+1}^{*} - \widebar{\rho}_{g,t+1}^{*},
    \; \forall g\in\mathcal{G},\; t\in\mathcal{T},
\end{equation}
with the boundary convention $\ushort{\rho}_{g,T+1}^{*}=\widebar{\rho}_{g,T+1}^{*}=0$. More importantly, the stationarity condition with respect to $P_{g,t}^G$ gives the main economic rule: at the optimum, marginal operation cost equals marginal benefit, net of the marginal damage from pollution
\vspace{-0.2cm}
\begin{multline}
\label{eq:welfare_maximisation}
\hspace{-0.4cm}
    \left.\frac{\partial \mathbf{C}_g^O}{\partial P_{g,t}^G}\right|_{P_{g,t}^G=P_{g,t}^{G*}}
    + \omega_{g,t}^{*}
    =
    \left.\frac{\partial \mathbf{U}_t}{\partial P_{g,t}^G}\right|_{P_{g,t}^G=P_{g,t}^{G*}}
    -
    \sum_{i\in\mathcal{I}}
    \left.\frac{d\mathbf{E}_i}{d Z_{i,t}}\right|_{Z_{i,t}=Z_{i,t}^{*}}
    \left.\frac{\partial Z_{i,t}}{\partial P_{g,t}^G}\right|_{P_{g,t}^G=P_{g,t}^{G*}}, \\
    \; \forall g\in\mathcal{G},\; t\in\mathcal{T}.
\vspace{-0.2cm}
\end{multline}
This condition makes clear why pollution matters. In particular, if marginal damages are positive, the planner prefers less output from generators that create those damages.

\vspace{-0.3cm}
\subsection{Competitive spot-market clearing}
\label{sec:competitive}
We now describe standard competitive spot-market clearing. The ISO clears the market subject to the same operational constraints, but it does not include the damage term $\sum_i \mathbf{E}_i(Z_{i,t})$ in the objective. Let $(\cdot)^\#$ denote the resulting competitive outcome. Under competitive clearing, the price at the location of generator $g$ in interval $t$ is the marginal value of consumption, i.e., the marginal aggregate utility evaluated at the competitive dispatch
\vspace{-0.2cm}
\begin{equation}
\hspace{-0.4cm}
\label{eq:price_consumer}
    \mathbf{\Theta}_{g,t}^{\#}\!\left(\widebar{P}_{g}^G\right)
    =
    \left.\frac{\partial \mathbf{U}_t}{\partial P_{g,t}^G}\right|_{P_{g,t}^G=P_{g,t}^{G\#}},
    \; \forall g\in\mathcal{G},\; t\in\mathcal{T}.
\end{equation}
We assume producers are price-takers in dispatch and must declare costs and technical limits truthfully. Given the market price, each generator chooses its dispatch to maximize operation profit
\vspace{-0.2cm}
\begin{equation}
\hspace{-0.4cm}
\label{eq:profit_original}
    \mathbf{\Pi}_{g}^{M\#}\!\left(\widebar{P}_{g}^G\right)
    =
    \max \sum_{t\in\mathcal{T}}
    \left[
        \mathbf{\Theta}_{g,t}^{\#}\!\left(\widebar{P}_{g}^G\right) P_{g,t}^G
        - \mathbf{C}_g^O(P_{g,t}^G)
    \right],
    \; \forall g\in\mathcal{G},
\vspace{-0.2cm}
\end{equation}
subject to \eqref{eq:relative_generation}--\eqref{eq:ramping_limits}. The corresponding KKT stationarity condition is
\vspace{-0.2cm}
\begin{equation}
\hspace{-0.4cm}
\label{eq:price_producer}
    \left.\frac{\partial \mathbf{C}_g^O}{\partial P_{g,t}^G}\right|_{P_{g,t}^G=P_{g,t}^{G\#}}
    + \omega_{g,t}^{\#}
    =
    \mathbf{\Theta}_{g,t}^{\#}\!\left(\widebar{P}_{g}^G\right),
    \; \forall g\in\mathcal{G},\; t\in\mathcal{T}.
\end{equation}

Comparing \eqref{eq:price_producer} with the planner’s condition \eqref{eq:welfare_maximisation}, we see the missing term. Competitive clearing does not subtract marginal pollution damages from marginal aggregate utility. When damages are positive, this tends to push the competitive outcome toward more output from polluting sources and, therefore, higher total pollution than in the social optimum.

\vspace{-0.2cm}
\subsection{Proposed tax scheme}
\label{sec:tax}
We now introduce a tax that corrects for the missing pollution term in competitive dispatch. The idea is simple: if pollution damages are a real social cost, then generators should face them as a private cost when choosing output. Let $\mathbf{\Phi}_{g,t}(P_{g,t}^G)$ be the tax charged to generator $g$ in interval $t$. Under the tax scheme, the dispatch price that clears the market is still given by marginal utility evaluated at the socially optimal dispatch
\vspace{-0.2cm}
\begin{equation}
\hspace{-0.4cm}
\label{eq:price_modified}
    \mathbf{\Theta}_{g,t}^{*}\!\left(\widebar{P}_{g}^G\right)
    =
    \left.\frac{\partial \mathbf{U}_t}{\partial P_{g,t}^G}\right|_{P_{g,t}^G=P_{g,t}^{G*}},
    \; \forall g\in\mathcal{G},\; t\in\mathcal{T}.
\end{equation}
With the tax, the generator $g$ maximizes profit net of operation costs and tax payments
\begin{equation}
\hspace{-0.4cm}
\label{eq:maximum_profit_tax}
    \mathbf{\Pi}_{g}^{M*}\!\left(\widebar{P}_{g}^G\right)
    =
    \max \sum_{t\in\mathcal{T}}
    \left[
        \mathbf{\Theta}_{g,t}^{*}\!\left(\widebar{P}_{g}^G\right) P_{g,t}^G
        - \mathbf{C}_g^O(P_{g,t}^G)
        - \mathbf{\Phi}_{g,t}(P_{g,t}^G)
    \right],
    \; \forall g\in\mathcal{G},
\end{equation}
subject to \eqref{eq:relative_generation}--\eqref{eq:ramping_limits}. The stationarity condition becomes
\vspace{-0.2cm}
\begin{equation}
\hspace{-0.4cm}
\label{eq:price_producer_modified}
    \left.\left(\frac{\partial \mathbf{C}_g^O}{\partial P_{g,t}^G}
    +\frac{\partial \mathbf{\Phi}_{g,t}}{\partial P_{g,t}^G}\right)\right|_{P_{g,t}^G=P_{g,t}^{G*}}
    + \omega_{g,t}^{*}
    =
    \mathbf{\Theta}_{g,t}^{*}\!\left(\widebar{P}_{g}^G\right),
    \; \forall g\in\mathcal{G},\; t\in\mathcal{T}.
\end{equation}

To make \eqref{eq:price_producer_modified} match the social optimum condition \eqref{eq:welfare_maximisation}, we choose the marginal tax to equal the marginal external damage
\vspace{-0.2cm}
\begin{equation}
\hspace{-0.4cm}
\label{eq:optimal_tax_condition}
    \frac{\partial \mathbf{\Phi}_{g,t}}{\partial P_{g,t}^G}
    =
    \sum_{i\in\mathcal{I}}
    \left.\frac{d\mathbf{E}_i}{d Z_{i,t}}\right|_{Z_{i,t}}
    \frac{\partial Z_{i,t}}{\partial P_{g,t}^G},
    \; \forall g\in\mathcal{G},\; t\in\mathcal{T}.
\end{equation}
Let $Z_{i,t}^{(-g)}$ denote total pollution at bus $i$ in interval $t$ when generator $g$ is fixed at zero output (while all other outputs are unchanged). By integrating \eqref{eq:optimal_tax_condition} from $0$ to $P_{g,t}^G$, we obtain
\vspace{-0.2cm}
\begin{equation}
\hspace{-0.4cm}
\label{eq:tax}
    \mathbf{\Phi}_{g,t}(P_{g,t}^G) - \mathbf{\Phi}_{g,t}(0)
    =
    \sum_{i\in\mathcal{I}}
    \left[
        \mathbf{E}_i(Z_{i,t}) - \mathbf{E}_i\!\left(Z_{i,t}^{(-g)}\right)
    \right],
    \; \forall g\in\mathcal{G},\; t\in\mathcal{T}.
\end{equation}

This tax is Pigouvian in that it charges each generator for the damages it adds at the margin~\citep{Sandmo2016}. Once the tax is applied, the generator’s private trade-off between revenue and cost includes pollution damages, and the profit-maximizing dispatch implements the socially optimal dispatch. In later sections, we use this corrected spot-market outcome as the basis for the investment analysis. Implementation requires generator-level emissions to be measurable through direct emissions monitoring or certified technology-specific emission factors combined with metered generation. The pollution-damage functions are assumed to be specified externally by the regulator or market operator.
\vspace{-0.3cm}
\section{Generation capacity investment in wholesale power markets} 
\label{sec:investment}
The previous section examined short-run dispatch and showed that a Pigouvian tax can align competitive dispatch with the social-welfare benchmark when pollution damages are internalized. We now consider long-run investment in generation capacity. Installed capacity affects future dispatch, network congestion, and spot-market prices. Consequently, producers may account for the effect of their capacity decisions on future market outcomes even when they behave as price-takers in the dispatch stage. We first formulate the socially optimal investment problem, then characterize strategic private investment, and finally derive a generator-specific subsidy that corrects the resulting investment incentive. Installed generation capacities and the associated investment decisions are assumed to be observable. The regulator is also assumed to announce and maintain the tax--subsidy rule, including its funding and settlement arrangements, over the relevant investment horizon. Strategic behavior in the investment stage is limited to producers accounting for the effect of their capacity choices on future spot-market dispatch and prices.

\vspace{-0.2cm}
\subsection{Socially optimal generation capacity investment}
Let the existing generation capacity of generator $g$ be $\widebar{P}_{g}^{G0}$, and let $\Delta \widebar{P}_{g}^G$ denote additional generation capacity investment. We write the non-negativity constraint as
\vspace{-0.0cm}
\begin{equation}
\label{eq:investment_limit}
    \Delta \widebar{P}_{g}^G \geq 0 \leftrightarrow \tau_g,\; \forall g\in\mathcal{G},
\vspace{-0.2cm}
\end{equation}
where $\tau_g \le 0$ is the associated multiplier. Total generation capacity is then
\vspace{-0.0cm}
\begin{equation}
\label{eq:incremental_capacity}
    \widebar{P}_{g}^G = \widebar{P}_{g}^{G0} + \Delta \widebar{P}_{g}^G,\; \forall g\in\mathcal{G}.
\end{equation}

Investment in generation capacity has cost $\mathbf{C}_{g}^I(\Delta \widebar{P}_{g}^G)$, which we assume is non-negative, non-decreasing, and convex. On the investment timescale, the social planner chooses capacities to maximize total social welfare minus investment costs
\vspace{-0.2cm}
\begin{equation}
\label{eq:investment_welfare_objective}
    \mathbf{W}^I
    =
    \max\;
    \mathbf{W}^M\!\left(\widebar{P}_{g}^G\right)
    -
    \sum_{g\in\mathcal{G}}
    \mathbf{C}_{g}^I\!\left(\Delta \widebar{P}_{g}^G\right),
\end{equation}
subject to \eqref{eq:investment_limit}--\eqref{eq:incremental_capacity}. We use $(\cdot)^*$ for the socially optimal investment solution. The first-order condition compares the marginal investment cost with the marginal welfare value of generation capacity
\vspace{-0.2cm}
\begin{equation}
\label{eq:welfare_maximisation_example}
    \left.\frac{\partial \mathbf{C}_{g}^I}{\partial \Delta \widebar{P}_{g}^G}\right|_{\Delta \widebar{P}_{g}^G=\Delta \widebar{P}_{g}^{G*}}
    + \tau_g^*
    =
    \left.\frac{\partial \mathbf{W}^M}{\partial \widebar{P}_{g}^G}\right|_{\widebar{P}_{g}^G=\widebar{P}_{g}^{G*}},
    \; \forall g\in\mathcal{G}.
\vspace{-0.2cm}
\end{equation}
This yields
\vspace{-0.2cm}
\begin{multline}
\hspace{-0.4cm}
\label{eq:welfare_maximisation_investment}
    \left. \left[ \sum_{t \in \mathcal{T}} \left. \left( \frac {\partial \mathbf{C}_{g}^O} {\partial P_{g,t}^G} + \sum_{i \in \mathcal{I}} \frac{\partial \mathbf{E}_i}{\partial Z_{i,t}} \frac{\partial Z_{i,t}}{\partial P_{g,t}^G} \right) \right \vert_{P_{g,t}^G = P_{g,t}^{G*}} \frac {\partial P_{g,t}^{G*}} {\partial \widebar{P}_{g}^G} + \frac {\partial \mathbf{C}_{g}^I} {\partial \Delta \widebar{P}_{g}^G} \right] \right \vert_{ \Delta \widebar{P}_{g}^G = \Delta \widebar{P}_{g}^{G*}} \\ + \tau_{g}^* = \sum_{t \in \mathcal{T}} \left. \mathbf{\Theta}_{g,t}^{*} \frac {\partial P_{g,t}^{G*}} {\partial \widebar{P}_{g}^G} \right \vert_{\Delta \widebar{P}_{g}^G = \Delta \widebar{P}_{g}^{G*}}.
\vspace{-0.2cm}
\end{multline}

\vspace{-0.2cm}
\subsection{Strategic generation capacity investment}
We now consider a producer that is a price-taker in dispatch but invests in generation capacity strategically because generation capacity affects future spot prices. We assume that the Pigouvian tax derived in Section~\ref{sec:tax} is applied, so dispatch is efficient conditional on installed capacities. However, the spot price $\mathbf{\Theta}_{g,t}^{*}$ still depends on capacities, and a producer may benefit from withholding generation capacity in order to affect prices. Let $\mathbf{\Pi}_{g}^{M*}(\widebar{P}_{g}^G)$ denote generator $g$'s operation profit in the taxed spot market, and let the producer choose $\Delta \widebar{P}_{g}^G$ to maximize profit net of investment costs
\vspace{-0.2cm}
\begin{equation}
\label{eq:maximum_profit}
    \mathbf{\Pi}_{g}^I
    =
    \max\;
    \mathbf{\Pi}_{g}^{M*}\!\left(\widebar{P}_{g}^G\right)
    -
    \mathbf{C}_{g}^I\!\left(\Delta \widebar{P}_{g}^G\right),
    \; \forall g\in\mathcal{G},
\end{equation}
subject to \eqref{eq:investment_limit}--\eqref{eq:incremental_capacity}. We use $(\cdot)^\#$ for the resulting strategic investment. The KKT condition for the strategic investment problem is
\vspace{-0.2cm}
\begin{equation}
\label{eq:profit_maximisation_example}
    \left.\frac{\partial \mathbf{C}_{g}^I}{\partial \Delta \widebar{P}_{g}^G}\right|_{\Delta \widebar{P}_{g}^G=\Delta \widebar{P}_{g}^{G\#}}
    + \tau_g^\#
    =
    \left.\frac{\partial \mathbf{\Pi}_{g}^{M*}}{\partial \widebar{P}_{g}^G}\right|_{\widebar{P}_{g}^G=\widebar{P}_{g}^{G\#}},
    \; \forall g\in\mathcal{G},
\vspace{-0.2cm}
\end{equation}
which yields
\vspace{-0.2cm}
\begin{multline}
\hspace{-0.4cm}
\label{eq:profit__maximising_condition_investment}
    \left. \left[ \sum_{t \in \mathcal{T}} \left. \left( \frac {\partial \mathbf{C}_{g}^O} {\partial P_{g,t}^G} + \sum_{i \in \mathcal{I}} \frac{\partial \mathbf{E}_i}{\partial Z_{i,t}} \frac{\partial Z_{i,t}}{\partial P_{g,t}^G} \right) \right \vert_{P_{g,t}^G = P_{g,t}^{G*}} \frac {\partial P_{g,t}^{G*}} {\partial \widebar{P}_{g}^G} + \frac {\partial \mathbf{C}_{g}^I} {\partial \Delta \widebar{P}_{g}^G} \right] \right \vert_{ \Delta \widebar{P}_{g}^G = \Delta \widebar{P}_{g}^{G\#}} \\ + \tau_{g}^\# = \sum_{t \in \mathcal{T}} \left. \left( \frac {\partial \mathbf{\Theta}_{g,t}^{*}} {\partial \widebar{P}_{g}^G} P_{g,t}^{G*} + \mathbf{\Theta}_{g,t}^{*} \frac {\partial P_{g,t}^{G*}} {\partial \widebar{P}_{g}^G} \right) \right \vert_{ \Delta \widebar{P}_{g}^G = \Delta \widebar{P}_{g}^{G\#}}, \; \forall g \in \mathcal{G}.
\vspace{-0.2cm}
\end{multline}

Comparing \eqref{eq:welfare_maximisation_example}--\eqref{eq:welfare_maximisation_investment} with \eqref{eq:profit_maximisation_example}--\eqref{eq:profit__maximising_condition_investment} shows a mismatch between the KKT conditions for socially optimal generation investment and strategic generation investment. The social planner values generation capacity only through the scarcity-value term. By contrast, the producer also values the effect of generation capacity on prices, which increases profit. Therefore, in general, $\Delta \widebar{P}_{g}^{G\#} \neq \Delta \widebar{P}_{g}^{G*}, \; \forall g\in\mathcal{G}$. Note that depending on the sign and magnitude of the price effect, the distortion may lead to either excessive or insufficient investment.

The comparison of the KKT conditions identifies the economic source of the difference between private and socially optimal investment. In the generation-capacity investment game, the KKT condition characterizes each producer's best response under the assumptions of the model. The term $P_{g,t}^{G*}\, \partial\mathbf{\Theta}_{g,t}^{*}/\partial\widebar{P}_{g}^{G}$ measures the effect of installed capacity on the market price received by the producer. This effect of capacity investment on market prices enters private profit but does not represent an additional benefit in the social-welfare problem. The subsidy derived in the next subsection corrects this specific investment incentive. This approach is consistent with a broader tradition in regulatory economics in which incentive mechanisms rules are designed so that a firm's profit-maximizing decisions are better aligned with regulatory or social-welfare objectives \citep{vogelsang1979regulatory, sappington1980strategic, hesamzadeh2018simple, khastieva2021optimal}. The contribution here is to apply this principle to the effect of capacity investment on market prices arising from strategic generation investment and to combine the resulting investment subsidy with a Pigouvian tax that corrects the pollution externality arising from short-term dispatch.

\vspace{-0.2cm}
\subsection{Proposed subsidy scheme}
\label{sec:subsidy}
We now design a subsidy that removes the price-effect profit, so that individual investment incentives match social incentives. Let $\mathbf{\Psi}_{g,t}(P_{g,t}^G)$ be the subsidy paid to generator $g$ in interval $t$ (a positive payment to the generator). Under the tax--subsidy scheme, the generator's operation payoff in each interval includes $+\mathbf{\Psi}_{g,t}$, so the investment condition becomes
\vspace{-0.2cm}
\begin{multline}
\hspace{-0.4cm}
\label{eq:maximum_profit_subsidy_condition}
    \left. \left[ \sum_{t \in \mathcal{T}} \left. \left( \frac {\partial \mathbf{C}_{g}^O} {\partial P_{g,t}^G} + \sum_{i \in \mathcal{I}}\frac{\partial \mathbf{E}_i}{\partial Z_{i,t}} \frac{\partial Z_{i,t}}{\partial P_{g,t}^G} - \frac {\partial \mathbf{\Psi}_{g,t}} {\partial P_{g,t}^G} \right) \right \vert_{P_{g,t}^G = P_{g,t}^{G*}} \frac {\partial P_{g,t}^{G*}} {\partial \widebar{P}_{g}^G} + \frac {\partial \mathbf{C}_{g}^I} {\partial \Delta \widebar{P}_{g}^G} \right] \right \vert_{ \Delta \widebar{P}_{g}^G = \Delta \widebar{P}_{g}^{G*}} \\ + \tau_{g}^\# = \sum_{t \in \mathcal{T}} \left. \left( \frac {\partial \mathbf{\Theta}_{g,t}^{*}} {\partial \widebar{P}_{g}^G} P_{g,t}^{G*} + \mathbf{\Theta}_{g,t}^{*} \frac {\partial P_{g,t}^{G*}} {\partial \widebar{P}_{g}^G} \right) \right \vert_{ \Delta \widebar{P}_{g}^G = \Delta \widebar{P}_{g}^{G*}}, \; \forall g \in \mathcal{G}.
\vspace{-0.2cm}
\end{multline}
To align this condition with the planner’s condition, it is sufficient to choose the subsidy so that it removes the price-effect term
\vspace{-0.2cm}
\begin{equation}
\label{eq:subsidy_marginal_condition}
    \left.\frac{\partial \mathbf{\Psi}_{g,t}}{\partial P_{g,t}^G}\right|_{P_{g,t}^G=P_{g,t}^{G*}}
    \frac{\partial P_{g,t}^{G*}}{\partial \widebar{P}_{g}^G}
    =
    -\left.\frac{\partial \mathbf{\Theta}_{g,t}^{*}}{\partial \widebar{P}_{g}^G}\right|_{\widebar{P}_{g}^G}
    P_{g,t}^{G*},
    \; \forall g\in\mathcal{G},\; t\in\mathcal{T}.
\end{equation}

Using the same idea as in the spot-market tax derivation, we can express the subsidy in a different form. Let $\mathbf{U}_t(\cdot)$ be the aggregate utility value function defined in Section~\ref{sec:spot_market}. Let $\mathbf{U}_{g,t}(\cdot)$ denote the same value function when the generator $g$ is fixed at zero output (and the market re-clears for the remaining units). Then
\vspace{-0.2cm}
\begin{multline}
\hspace{-0.4cm}
\label{eq:subsidy}
    \mathbf{\Psi}_{g,t}(P_{g,t}^G) - \mathbf{\Psi}_{g,t}(0)
    =
    \mathbf{U}_t(P_{g,t}^G)
    -
    \mathbf{U}_{g,t}(P_{g,t}^G)
    -
    \mathbf{\Theta}_{g,t}^{*}(P_{g,t}^G)\, P_{g,t}^{G}, 
    \; \forall g\in\mathcal{G},\; t\in\mathcal{T}.
\vspace{-0.2cm}
\end{multline}

The difference between aggregate utility under the original market-clearing outcome and aggregate utility when generator $g$ is fixed at zero output represents generator $g$'s contribution to aggregate utility. Subtracting the generator's market revenue from this contribution gives generator $g$'s contribution to consumer surplus. The investment-support component in \eqref{eq:subsidy} is constructed from this generator-specific consumer-surplus contribution. Up to the constant $\mathbf{\Psi}_{g,t}(0)$, the subsidy therefore equals the generator's contribution to consumer surplus. When the Pigouvian tax from Section~\ref{sec:tax} and the subsidy in \eqref{eq:subsidy} are applied jointly, the marginal conditions of the generator's profit-maximization problem coincide with those of the social-welfare problem. Under the stated assumptions, decentralized profit maximization therefore implements socially optimal dispatch and generation-capacity investment. 

Calculation of the subsidy requires a generator-specific counterfactual market re-clearing. In this counterfactual analysis, generator $g$ is fixed at zero output, while submitted bids, demand, network constraints, intertemporal constraints, and other system conditions are held fixed. The remaining generation and demand decisions are then re-optimized. Calculating the generator-specific aggregate-utility contribution requires one additional network-constrained market-clearing problem for each generator whenever the subsidy is evaluated. Although these counterfactual problems have similar structures and can be solved in parallel, the computational burden may be substantially expensive in large and multi-period systems. Developing decomposition, approximation, or screening methods for large-scale implementation is outside the scope of this paper.

\vspace{-0.3cm}
\section{Properties of the proposed scheme}
\label{sec:properties}
This section summarizes key properties of the proposed tax--subsidy scheme based on the tax and subsidy derived in Subsections~\ref{sec:tax} and~\ref{sec:subsidy}. The values $\left.\mathbf{\Phi}_{g,t}(P_{g,t}^G)\right|_{P_{g,t}^G=0}$ and $\left.\mathbf{\Psi}_{g,t}(P_{g,t}^G)\right|_{P_{g,t}^G=0}$ remain as design freedoms. Because these terms are independent of output and investment, they do not change marginal dispatch or investment incentives. They nevertheless affect producer participation and the budget of the mechanism. The two instruments address distinct sources of inefficiency. Comparing the competitive dispatch conditions with the social planner's conditions shows that the dispatch problem omits the marginal pollution damage caused by each generator; the proposed Pigouvian tax introduces this missing term. Comparing the private and socially optimal investment conditions identifies the effect of capacity investment on market prices in the producer's long-run decision; the proposed investment support is constructed to correct this term. Under the stated assumptions, these corrections align the corresponding private optimality conditions with the social-welfare benchmark.

\begin{itemize}
    \item \textbf{Individual rationality (with suitable constants).}
    If the participation of generator $g$ does not reduce social welfare in interval $t$, then the scheme can be made individually rational by choosing the constants at $P_{g,t}^G=0$ so that the generator’s payoff is at least as large as its payoff when it does not participate. A sufficient condition is that the constants satisfy
    \vspace{-0.2cm}
    \begin{multline}
    \hspace{-0.4cm}
    \label{eq:individually_rational}
        \left.\left(\mathbf{\Psi}_{g,t}-\mathbf{\Phi}_{g,t}\right)\right|_{P_{g,t}^G=0}
        \ge
        \mathbf{U}_{g,t}\!\left(P_{g,t}^G\right) - \mathbf{U}_{t}\!\left(P_{g,t}^G\right)
        + \mathbf{C}_g^O\!\left(P_{g,t}^G\right) \\
        + \sum_{i\in\mathcal{I}}
        \left(
        \mathbf{E}_i\!\left(Z_{i,t}\right) - \mathbf{E}_i\!\left(Z_{i,t}^{(-g)}\right)
        \right),
        \; \forall g\in\mathcal{G},
    \vspace{-0.2cm}
    \end{multline}
    where $\mathbf{U}_{g,t}(\cdot)$ is the aggregate utility when generator $g$ is fixed at zero output and the market re-clears, and $Z_{i,t}^{(-g)}$ is total pollution at bus $i$ when generator $g$ is fixed at zero output. In practice, the constants can be chosen based on the equilibrium outcome (or over the full horizon) rather than for every possible value of $P_{g,t}^G$. Specifically, let $\overline{\Pi}_{g}$ denote producer $g$'s profit when it does not participate in the proposed tax--subsidy mechanism, and let $\Pi_{g}^{\mathrm{mech},0}$ denote its equilibrium profit under the mechanism before the constant transfer. The minimum net constant transfer satisfying the participation constraint is $\max\left\{0,\overline{\Pi}_{g} -\Pi_{g}^{\mathrm{mech},0}\right\}$. This transfer may be allocated across intervals according to a predetermined settlement rule. Because it is fixed independently of subsequent output and investment decisions, it does not alter the producer's marginal incentives. It must, however, be included in the financing requirement of the mechanism.

    \item \textbf{Information and computational requirements.} 
    The proposed mechanism is neither information-free nor computation-free. It requires the information ordinarily used in market clearing, verified emissions and technical data, a regulator-specified pollution-damage function, and generator-specific counterfactual market re-clearing. Its narrower informational advantage is that it does not require a new producer-specific report of the complete private cost function beyond the submissions already used in market clearing and settlement.

    \item \textbf{Behavioral scope of the alignment result.} 
    The exact alignment result is conditional on competitive dispatch and truthful or independently verifiable operational information. If producers exercise residual dispatch-side market power through strategic bidding, economic or physical withholding, or the misreporting of technical constraints, additional distortions arise, and the proposed tax and subsidy alone need not implement the social-welfare benchmark. The analysis of a mechanism that jointly addresses operational and investment market power is left for future research.

     \item \textbf{Non-discriminatory through common rules for the constant terms.} 
     The variable parts of the tax and subsidy are based on generator-specific marginal contributions, but the same calculation rule applies to all generators. If the regulator also selects $\left.\mathbf{\Phi}_{g,t}\right|_{P_{g,t}^G=0}$ and $\left.\mathbf{\Psi}_{g,t}\right|_{P_{g,t}^G=0}$ according to the same rule for similarly situated generators, such as generators within the same technology category, the mechanism treats those generators symmetrically.

    \item \textbf{Incentives aligned with social-welfare maximization.}
    By construction, the Pigouvian tax makes each generator face its marginal pollution damage as a private cost, so profit-maximizing dispatch coincides with socially optimal dispatch. The subsidy corrects the incentive to distort generation-capacity investment through the effect of installed capacity on future spot prices. Therefore, under the assumptions of competitive dispatch, truthful or independently verifiable technical information, observable capacity decisions, and verifiable emissions, generators' profit maximization under the mechanism implements the socially optimal outcome.
    
    \item \textbf{Budget balance, financing, and regulatory commitment.}
    The proposed mechanism is not budget balanced in general. Budget balance would require aggregate pollution-tax receipts to equal aggregate investment-subsidy payments, including any constant transfers used to satisfy participation constraints. This equality is not imposed by the mechanism and need not hold at the resulting equilibrium. Budget balance is therefore a first-order implementation condition rather than a secondary property. The efficiency result should be distinguished from the separate institutional question of how the resulting funding deficit or surplus is settled.

    Pollution-tax receipts provide a natural initial source for financing the investment subsidies, consistent with the broader concept of environmental-revenue recycling \citep{gavard2022recycling}. If subsidy and constant-transfer payments exceed tax receipts, however, the remaining deficit must be financed from an external source, such as general public funds or a separately designed system charge. If tax receipts exceed total payments, the surplus must be assigned through an appropriate surplus-allocation rule, such as transfer to the reserve fund or the public budget. The exact alignment result is preserved only when the residual financing or surplus-allocation rule does not alter the marginal dispatch and investment incentives derived above. A distortionary financing instrument may change demand, dispatch, investment, and distributional outcomes and must therefore be evaluated jointly with the proposed mechanism.

    Because generation investment is a long-run decision, implementation also requires credible regulatory commitment to the announced tax, subsidy, funding, and settlement rules over the relevant investment horizon. Policy and regulatory uncertainty can weaken generation-investment incentives \citep{brunekreeft2005policy}, while the use and distribution of environmental revenues can affect political acceptability \citep{klenert2018making}. The present analysis assumes such commitment and does not establish the political sustainability or distributional acceptability of a particular financing arrangement. These issues, together with the design of a budget-balanced or otherwise credibly funded extension, are left for future research.

    \item \textbf{Reduced dependence of producer profits on spot-price rules.}
    A generator's payoff under the mechanism can be expressed in terms of aggregate utility, operating costs, pollution damages, and the constant transfer terms. Producer remuneration therefore does not rely exclusively on high spot prices. This property reduces dependence on scarcity pricing and may make the mechanism compatible with price caps, provided that the price-cap rule is incorporated consistently into market clearing and does not prevent implementation of the targeted dispatch and investment outcome.
\end{itemize}
\vspace{-0.3cm}
\section{Analytical Example}
\label{sec:example}
This section illustrates the main idea of the paper using a small analytical example. We first consider the spot market and compare (i) the socially optimal dispatch that accounts for pollution damages and (ii) the competitive dispatch that ignores them. We then move to the investment stage and show why a clean generator may underinvest when generation capacity affects prices. Finally, we show how the subsidy restores the socially optimal investment.

We study a single-bus system with $\mathcal{I}=\{1\}$, two producers $\mathcal{P}=\{1,2\}$, one generator for each producer $\mathcal{G}=\{1,2\}$, and one demand $\mathcal{D}=\{1\}$. Dispatch takes place over three intervals $\mathcal{T}=\{1,2,3\}$. Since there is only one bus, there are no transmission limits.

To simplify the constraints, we set $A_{g,t}^G=1$ and ignore ramping. Thus, generation is limited only by capacity, i.e., $0 \le P_{g,t}^G \le \widebar{P}_{g}^{G0}$, $\forall g\in\{1,2\}$, $\forall t\in\{1,2,3\}$. Demand must satisfy the power balance, $P_{1,t}^{D}=P_{1,t}^{G}+P_{2,t}^{G}$, $\forall t\in\{1,2,3\}$, and consumer utility is $\widetilde{\mathbf{U}}_{1,t}(P_{1,t}^{D}) = c_t P_{1,t}^{D} - \frac{(P_{1,t}^{D})^2}{2}$, where $c_1=6$, $c_2=12$, and $c_3=20$. Table~\ref{tab:data} reports operation costs, pollution, and existing capacities. Generator 1 is cheaper but has high pollution; generator 2 is more expensive but cleaner. We assume damages equal total pollution on the bus, i.e., $\mathbf{E}_1(Z_{1,t}) = Z_{1,t}$, where $Z_{1,t}=X_{1,1,t}+X_{2,1,t}$. Since generator 2 has zero pollution, we have $Z_{1,t}=4P_{1,t}^G$.

\begin{table}[t]
    \fontsize{10pt}{12pt}
    \caption{Operation cost, pollution, existing generation capacity, and investment cost}
    \label{tab:data}
    \vspace{-0.35cm}
    \centering
    \begin{tabular}{c||c||c||c||c}
        $g$ & $\mathbf{C}_g^O(P_{g,t}^G)$ & $\mathbf{X}_{g,1}(P_{g,t}^G)$ & $\widebar{P}_{g}^{G0}$ & $\mathbf{C}_{g}^I(\Delta \widebar{P}_{g}^G)$ \\
        \hline
        1 & $2 P_{1,t}^G$ & $4 P_{1,t}^G$ & 4 & $9\,\Delta \widebar{P}_{1}^G$ \\
        2 & $4 P_{2,t}^G$ & $0$            & 3 & $9\,\Delta \widebar{P}_{2}^G$ \\
    \end{tabular}
    \vspace{-0.35cm}
\end{table}

\begin{figure*}[t]
    \centering
    \begin{subfigure}[b]{0.27\textwidth}
        \centering
        \includegraphics[width=\textwidth]{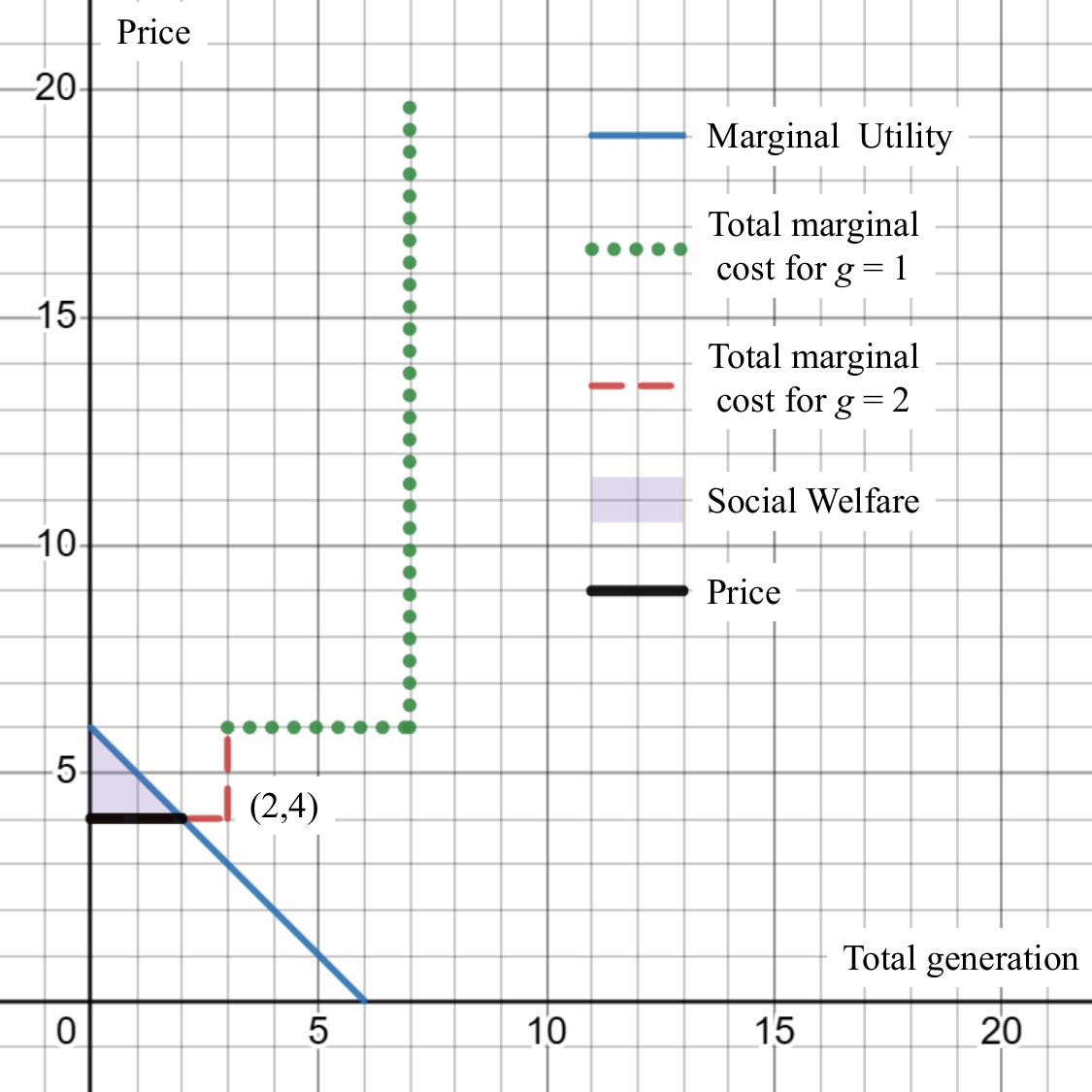}
        \caption{Interval $t=1$}
    \end{subfigure}
    \hfill
    \begin{subfigure}[b]{0.27\textwidth}
        \centering
        \includegraphics[width=\textwidth]{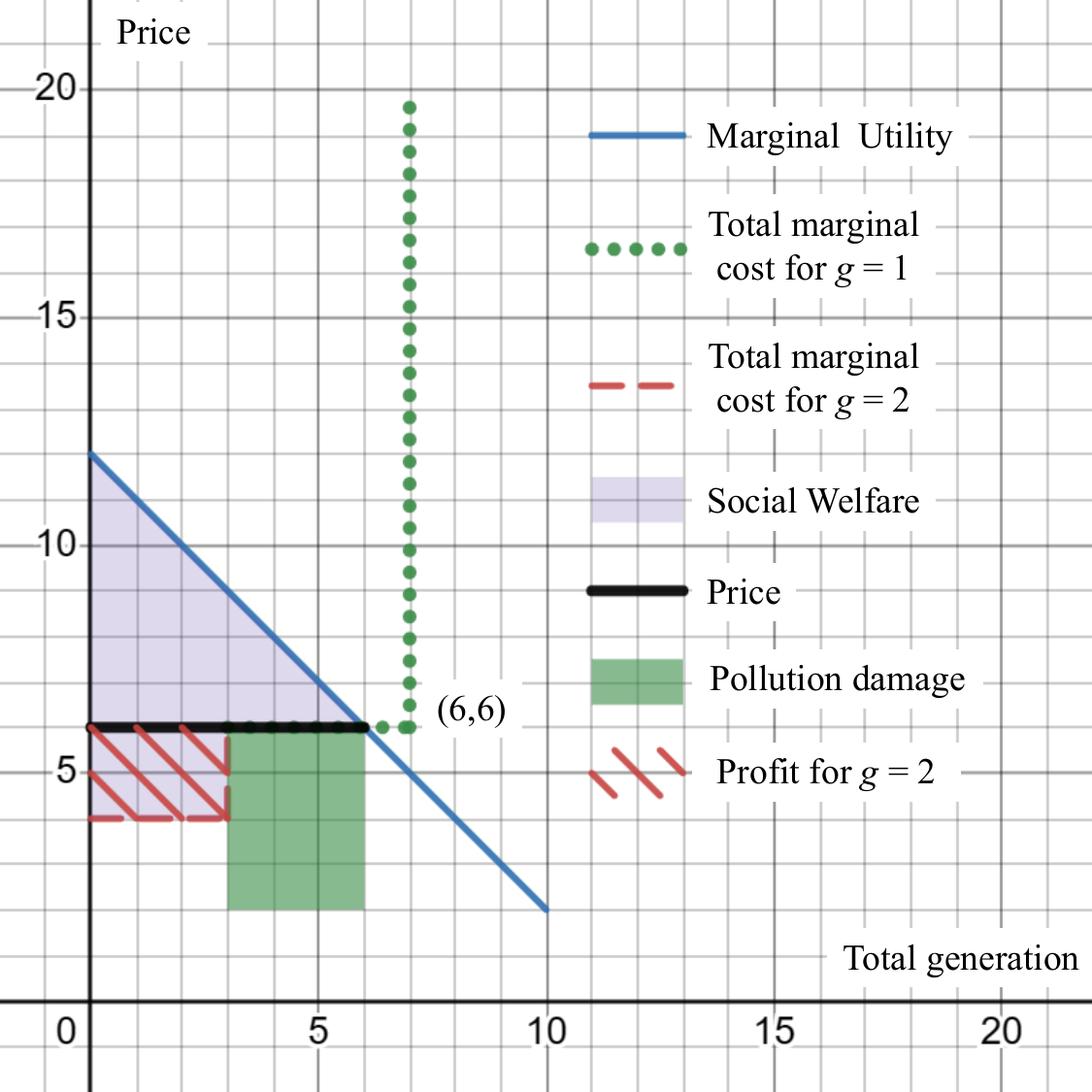}
        \caption{Interval $t=2$}
    \end{subfigure}
    \hfill
    \begin{subfigure}[b]{0.27\textwidth}
        \centering
        \includegraphics[width=\textwidth]{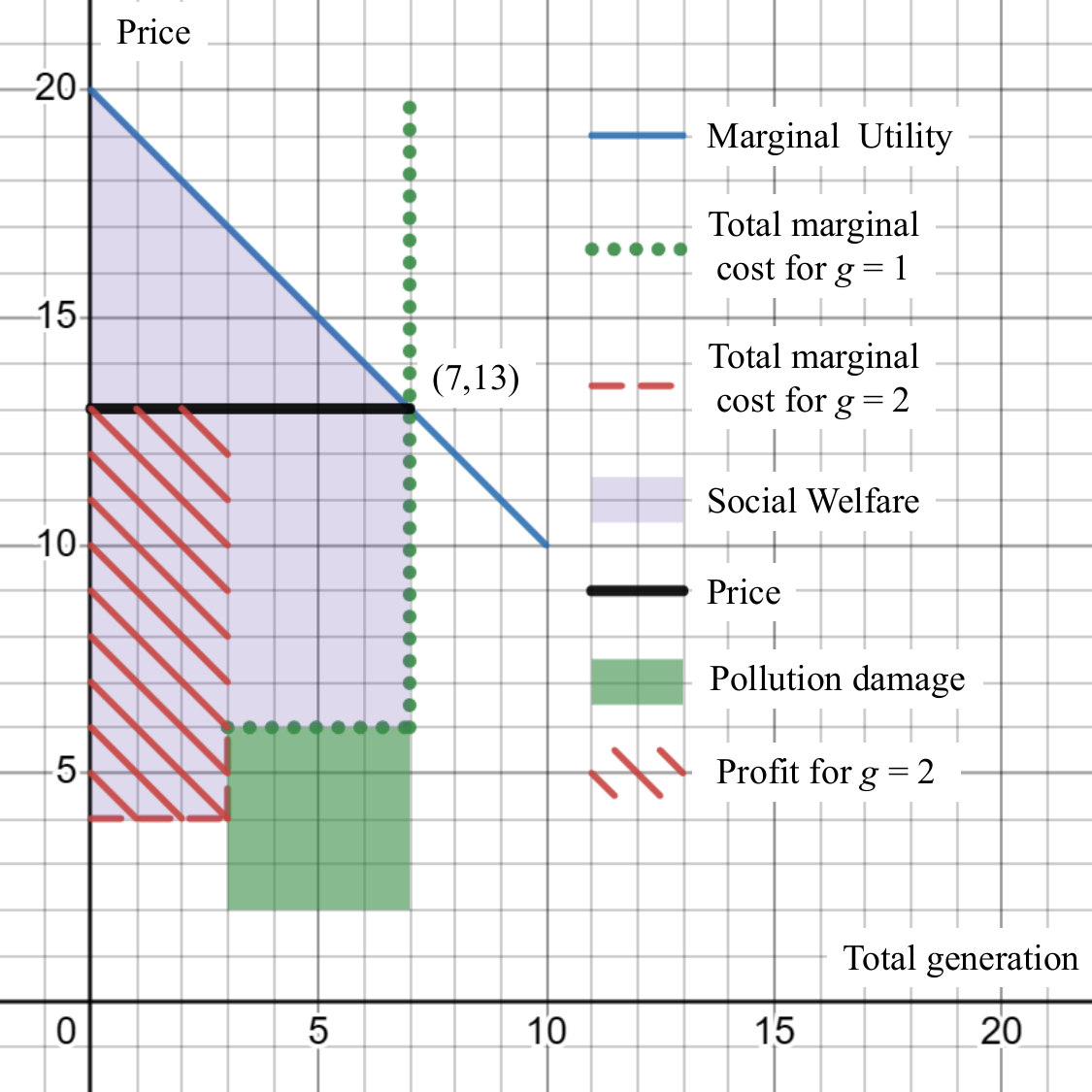}
        \caption{Interval $t=3$}
    \end{subfigure}
    \caption{Socially optimal spot-market dispatch (pollution damages included)}
    \label{fig:optimal}
    \vspace{-0.35cm}
\end{figure*}

In the social optimum, dispatch compares marginal benefit (marginal utility) with marginal social cost. Marginal utility is
$\frac{d\widetilde{\mathbf{U}}_{1,t}}{dP_{1,t}^{D}} = c_t - P_{1,t}^{D}$. The social marginal cost of generator 2 is $4$. For generator 1, it is $2$ plus the marginal damage $4$, so it is $6$. This means the planner uses the clean unit first and uses the polluting unit only when needed. Figure~\ref{fig:optimal} shows this for each interval, and Table~\ref{tab:results} summarizes the outcomes.

\begin{figure*}[t]
    \centering
    \begin{subfigure}[b]{0.27\textwidth}
        \centering
        \includegraphics[width=\textwidth]{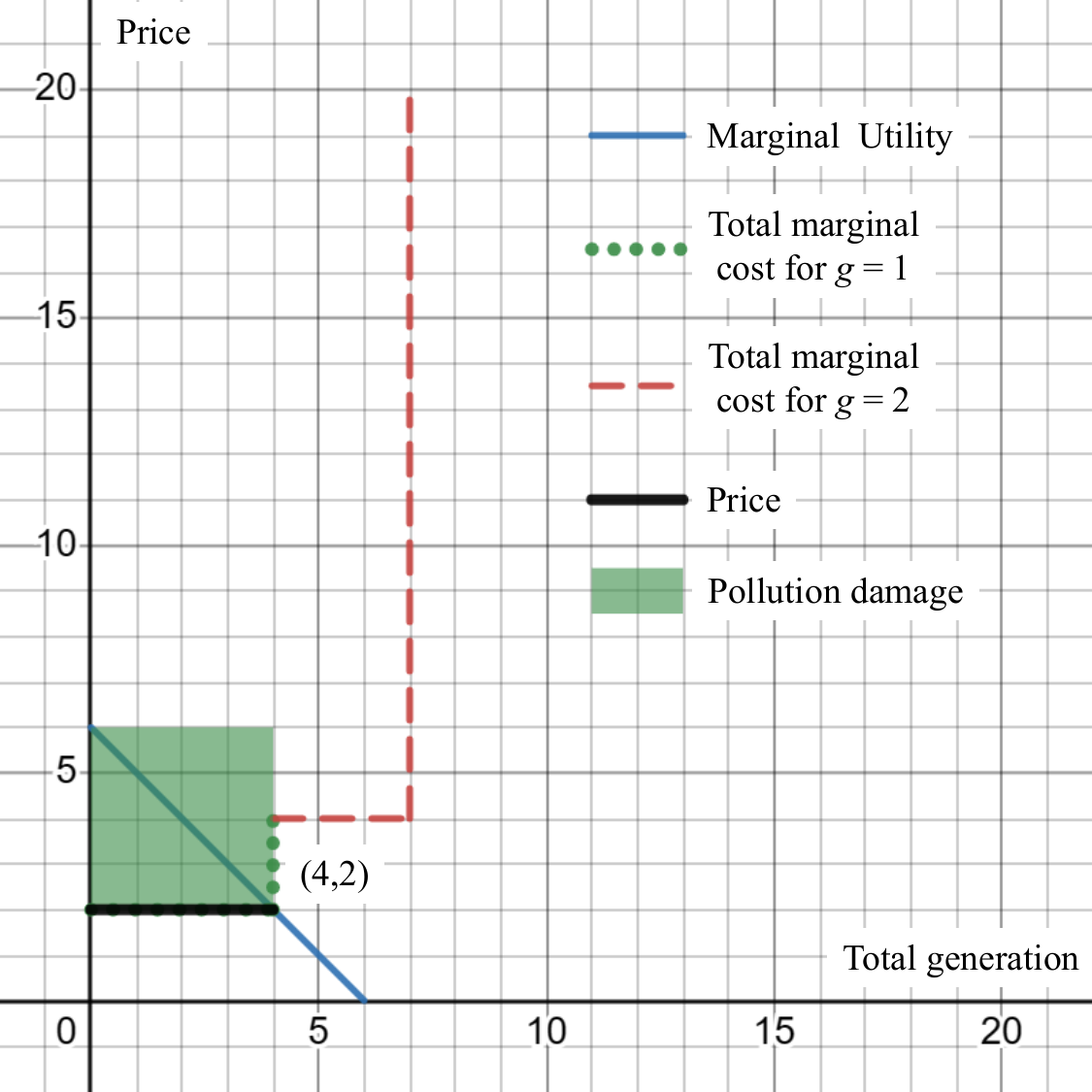}
        \caption{Interval $t=1$}
    \end{subfigure}
    \hfill
    \begin{subfigure}[b]{0.27\textwidth}
        \centering
        \includegraphics[width=\textwidth]{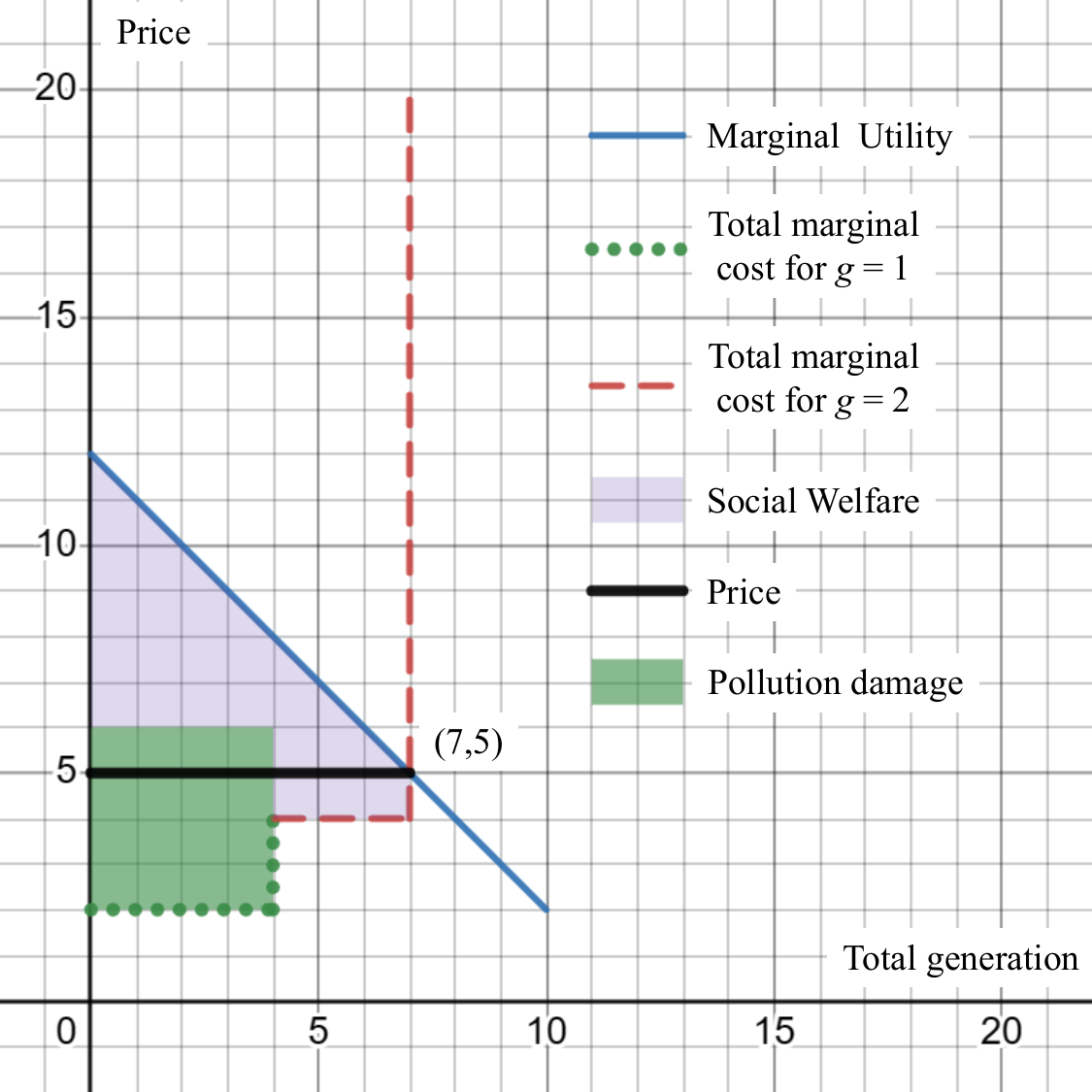}
        \caption{Interval $t=2$}
    \end{subfigure}
    \hfill
    \begin{subfigure}[b]{0.27\textwidth}
        \centering
        \includegraphics[width=\textwidth]{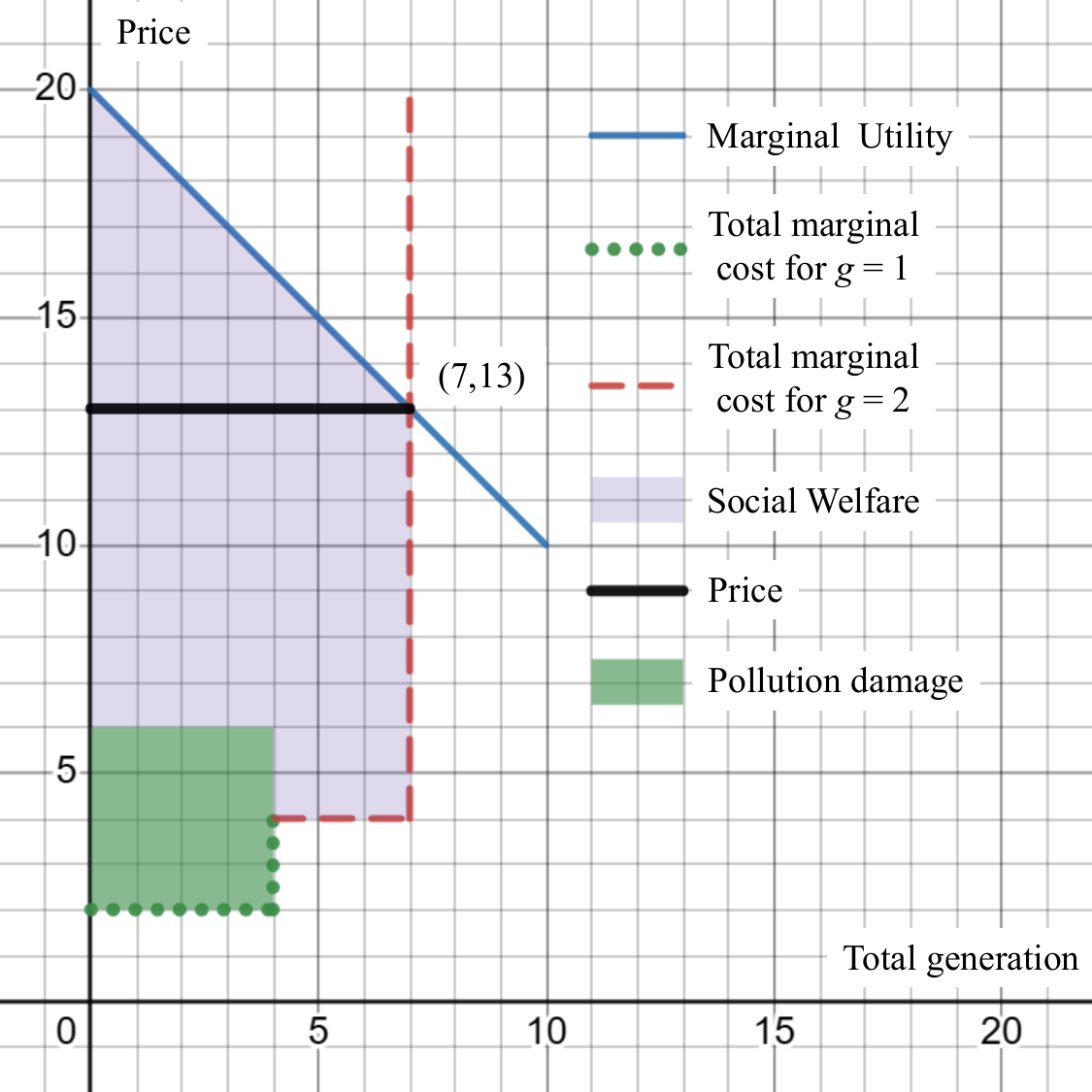}
        \caption{Interval $t=3$}
    \end{subfigure}
    \caption{Competitive spot-market clearing (pollution damages ignored)}
    \label{fig:competitive}
    \vspace{-0.25cm}
\end{figure*}

Under competitive clearing, we ignore $\mathbf{E}_1(Z_{1,t})$. Then marginal costs are $2$ for generator 1 and $4$ for generator 2, so the dispatch order reverses: the market uses the polluting unit first. Figure~\ref{fig:competitive} shows the resulting dispatch. In interval $t=1$, this leads to high pollution and negative welfare once damages are taken into account (Table~\ref{tab:results}).

\begin{table*}[t]
    \centering
    \fontsize{10pt}{12pt}
    \caption{Optimal vs.\ competitive dispatch (sw = welfare; tax shown for the polluting unit)}
    \label{tab:results}
    \vspace{-0.35cm}
    \resizebox{1.00\columnwidth}{!}{
    \begin{tabular}{c||c||c c c||c||c||c||c}
        & $t$ & $P_{1,t}^{G}$ & $P_{2,t}^{G}$ & $P_{1,t}^{D}$ &
        $\Theta_{1,t}, \Theta_{2,t}$ &
        $\mathbf{E}_1(Z_{1,t})$ & sw &
        $\mathbf{\Phi}_{1,t}(P_{1,t}^G)-\mathbf{\Phi}_{1,t}(0)$ \\
        \hline
        & 1 & 0 & 2 & 2 & 4  & 0  & 2    & 0 \\
        Optimal & 2 & 3 & 3 & 6 & 6  & 12 & 24   & 12 \\
        & 3 & 4 & 3 & 7 & 13 & 16 & 79.5 & 16 \\
        \hline
        & 1 & 4 & 0 & 4 & 2  & 16 & -8   & -- \\
        Competitive & 2 & 4 & 3 & 7 & 5  & 16 & 23.5 & -- \\
        & 3 & 4 & 3 & 7 & 13 & 16 & 79.5 & -- \\
    \end{tabular}
    }
    \vspace{-0.55cm}
\end{table*}

The Pigouvian tax from Subsection~\ref{sec:tax} charges each unit for the damage it adds. Here, only generator 1 creates pollution, so the variable part of its tax is $\mathbf{\Phi}_{1,t}(P_{1,t}^G)-\mathbf{\Phi}_{1,t}(0) = \mathbf{E}_1(4P_{1,t}^G)-\mathbf{E}_1(0)=4P_{1,t}^G$,
which matches the last column of Table~\ref{tab:results}. With this tax, the private cost of generator 1 includes the damage term, and market dispatch coincides with the social optimum.

\begin{figure*}[t]
    \centering
    \begin{subfigure}[b]{0.27\textwidth}
        \centering
        \includegraphics[width=\textwidth]{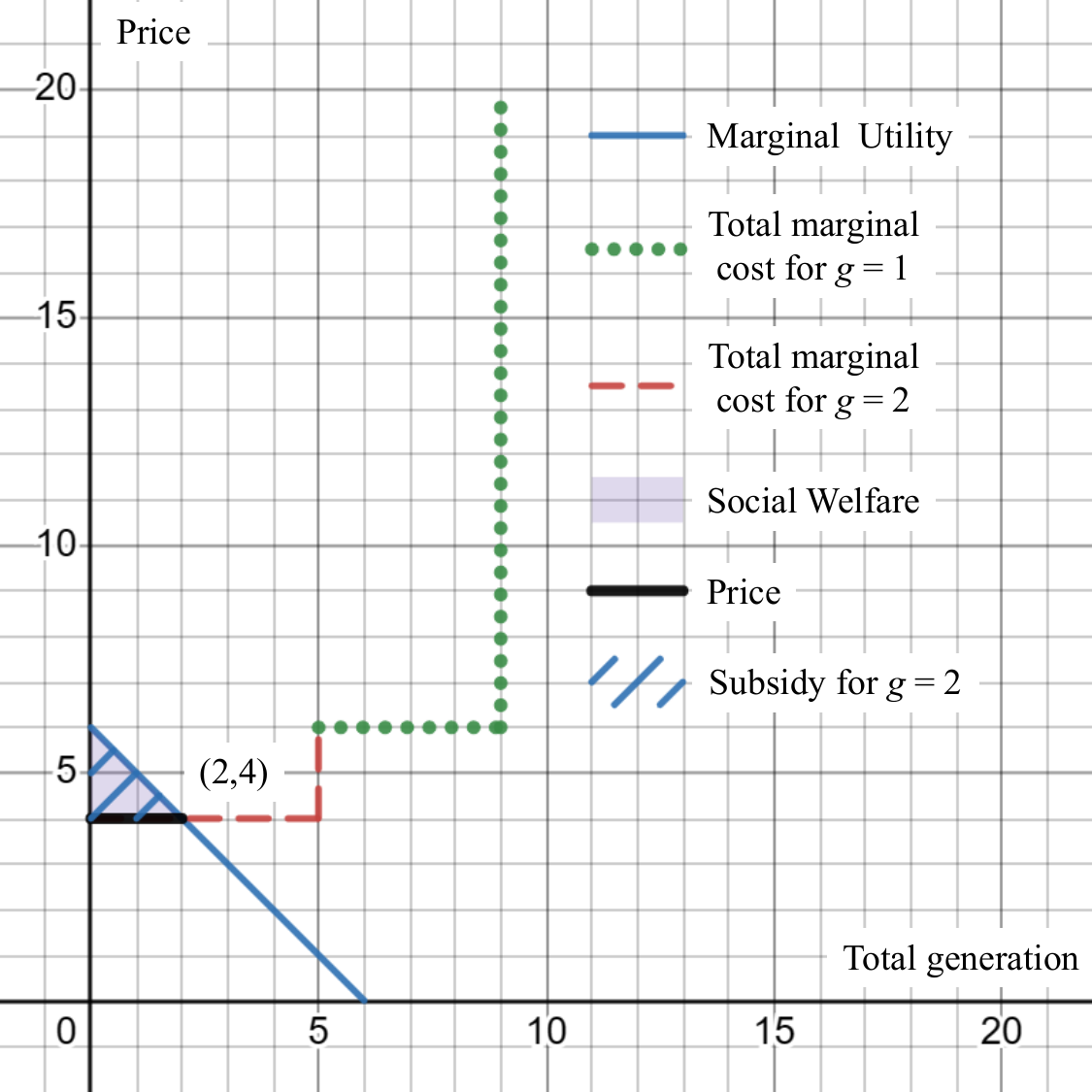}
        \caption{Interval $t=1$}
    \end{subfigure}
    \hfill
    \begin{subfigure}[b]{0.27\textwidth}
        \centering
        \includegraphics[width=\textwidth]{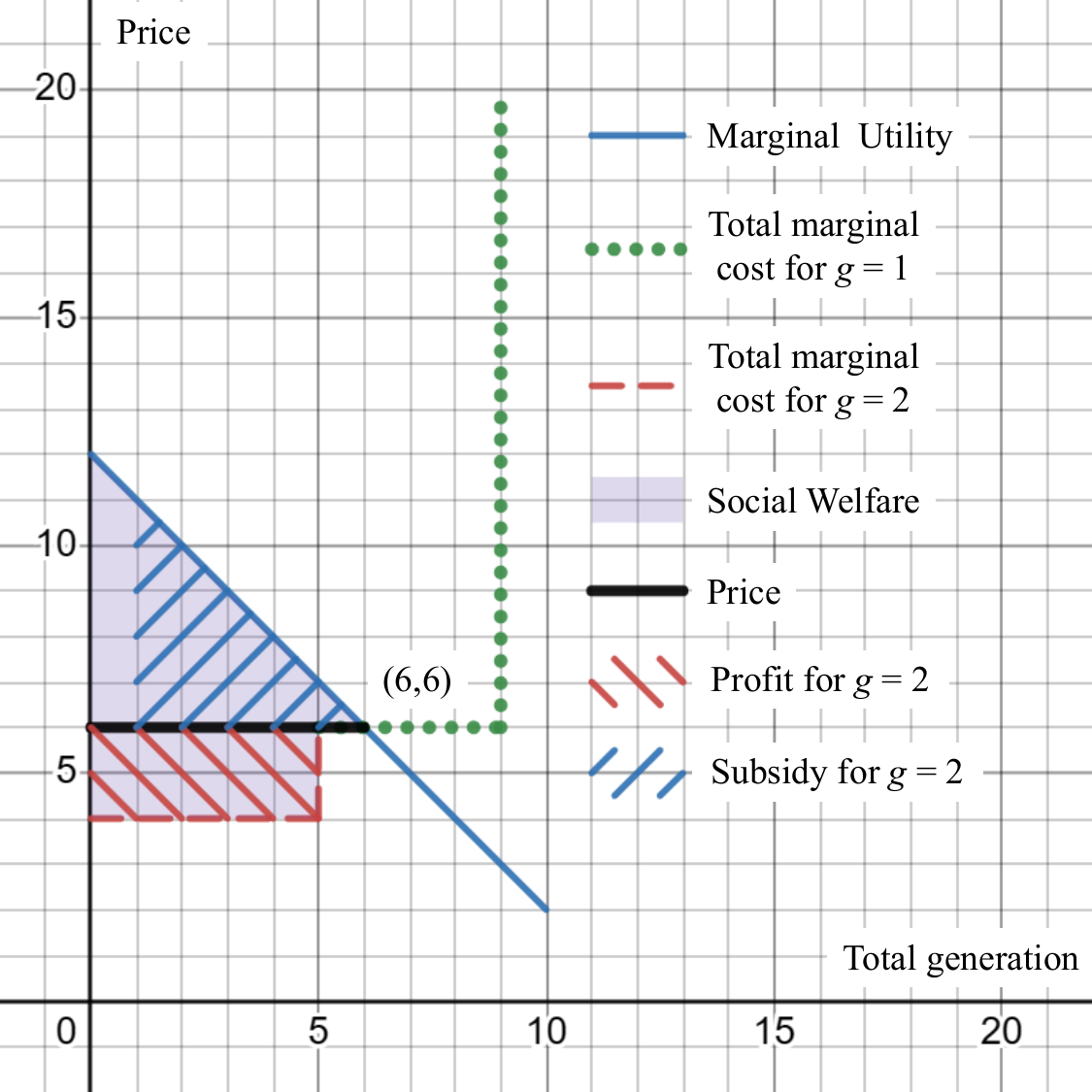}
        \caption{Interval $t=2$}
    \end{subfigure}
    \hfill
    \begin{subfigure}[b]{0.27\textwidth}
        \centering
        \includegraphics[width=\textwidth]{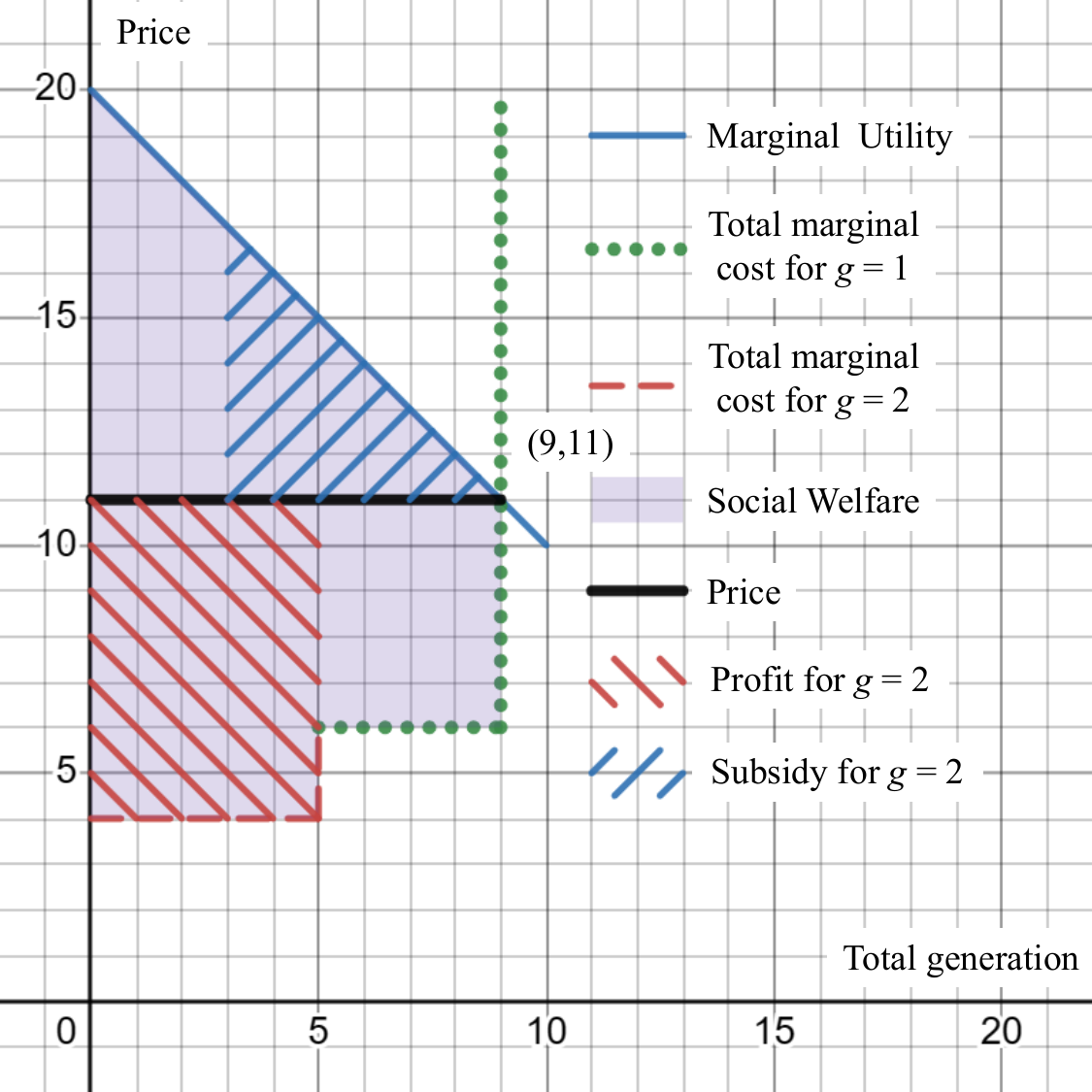}
        \caption{Interval $t=3$}
    \end{subfigure}
    \caption{Spot-market outcomes under the socially optimal generation capacity $k=\widebar{P}_2^G$}
    \label{fig:optimal_investment}
    \vspace{-0.25cm}
\end{figure*}

Now, consider an investment problem with existing capacities $\widebar{P}_{1}^{G0}=4$ and $\widebar{P}_{2}^{G0}=3$, and with linear investment costs, i.e., $\mathbf{C}_{g}^I(\Delta \widebar{P}_{g}^G)=9\,\Delta \widebar{P}_{g}^G$. Since generator 2 has the lower social marginal cost (it is clean), the planner expands only generator 2 and keeps $\widebar{P}_{1}^{G}=\widebar{P}_{1}^{G0}=4$. Let $k:= \widebar{P}_{2}^{G}$ denote generator 2's post-investment capacity. For the range of $k$ considered here, the socially optimal dispatch and price are shown in Table~\ref{tab:generation}, and Figure~\ref{fig:optimal_investment} illustrates the three intervals considered.

\begin{table}[t]
    \centering
    \fontsize{10pt}{12pt}
    \caption{Dispatch and price as functions of generator 2 generation capacity $k=\widebar{P}_{2}^{G}$}
    \label{tab:generation}
    \vspace{-0.35cm}
    \begin{tabular}{c||c c c||c}
         & $P_{1,t}^{G}$ & $P_{2,t}^{G}$ & $P_{1,t}^{D}$ &
         $\mathbf{\Theta}_{g,t}^*\!\left(\widebar{P}_{1}^{G0}, k\right),\; \forall g\in\{1,2\}$\\
        \hline
        $t = 1$ & 0 & 2 & 2 & 4 \\
        $t = 2$ & $6 - k$ & $k$ & 6 & 6 \\
        $t = 3$ & 4 & $k$ & $4 + k$ & $16 - k$ \\
    \end{tabular}
    \vspace{-0.55cm}
\end{table}

Using Table~\ref{tab:generation}, spot-market welfare (with the Pigouvian tax in place) can be written as a function of $k$, i.e., $\mathbf{W}^I(\widebar{P}_{1}^{G0},k)= 68 + 14k - \frac{k^2}{2}$. This implies that the planner chooses $k^*=5$, so the socially optimal investment is $\Delta \widebar{P}_{2}^{G*}=2$. The resulting dispatch, prices, welfare, and transfers are shown in Table~\ref{tab:results_investment}.

Now consider producer 2 investing to maximize its own profit under the Pigouvian tax but without the subsidy. From Table~\ref{tab:generation}, producer 2's total operation profit in the spot market (before investment cost) is $\mathbf{\Pi}_{2}^M(\widebar{P}_{1}^{G0},k)=14k-k^2$. Using \eqref{eq:welfare_maximisation_example} in this case, the best feasible choice is $k^\#=3$, which implies no investment: $\Delta \widebar{P}_{2}^{G\#}=0$. The reason is simple: more clean generation capacity reduces the price in interval $t=3$, thereby reducing the producer's revenue.

Finally, we apply the subsidy from Subsection~\ref{sec:subsidy}. The subsidy pays generator 2 according to its contribution to consumer surplus. In this example, the variable part of the subsidy equals $2\ \$$ at $t=1$ and $12.5\ \$$ at $t=2$ and $t=3$ under the optimal generation capacity $k^*=5$ MW. With this subsidy, the private gain from investing matches the social gain, and the producer chooses the socially optimal investment.

\begin{table*}[t]
    \fontsize{10pt}{12pt}
    \centering
    \caption{Optimal vs.\ strategic investment (ps$_2$ = profit of producer 2 in the spot market after tax; sw = spot-market welfare)}
    \label{tab:results_investment}
    \vspace{-0.35cm}
    \resizebox{1.00\columnwidth}{!}{
    \begin{tabular}{c||c||c c c||c||c||c||c||c} 
         & $t$ & $P_{1,t}^G$ & $P_{2,t}^G$ & $P_{1,t}^D$ &
         $\mathbf{\Theta}_{g,t}^*\!\left(\widebar{P}_{1}^{G0}, \widebar{P}_{2}^{G}\right)$ &
         $\mathbf{\Phi}_{1,t}(P_{1,t}^G)-\mathbf{\Phi}_{1,t}(0)$ &
         ps$_2$ & sw &
         $\mathbf{\Psi}_{2,t}(P_{2,t}^G)-\mathbf{\Psi}_{2,t}(0)$ \\
        \hline
        &1&0&2&2&4&0&0&2&2\\
        Optimal&2&1&5&6&6&4&10&28&12.5\\
        &3&4&5&9&11&16&10&95.5&12.5\\
        \hline
        &1&0&2&2&4&0&0&2&--\\
        Strategic&2&3&3&6&6&12&12&24&--\\
        &3&4&3&7&13&16&16&79.5&--\\
    \end{tabular}
    }
    \vspace{-0.25cm}
\end{table*}
\vspace{-0.3cm}
\section{Numerical Experiments}
\label{sec:experiments}

This section evaluates the proposed tax--subsidy scheme using the IEEE 24-bus test system, which contains 33 generators and 38 transmission lines. The numerical study examines how the proposed mechanism affects (i) spot-market outcomes when pollution damages are internalized and (ii) generation-capacity investment when producers account for the effect of capacity on future prices. The 33 generators are grouped into ten producers. Each producer owns a mix of technologies from the following set: renewable, hydropower, nuclear, coal, fuel oil, and gas. Before investment, the total installed generation capacity is 3405~MW, and the total load in the first time step is 2850~MW. The simulation covers 20 annual time steps.

To maintain tractability and consistency with the formulation in~\ref{sec:Appendix}, the utility functions of demands and the operating-cost functions of generators are represented by piecewise linear functions with ten segments. Pollution damages are represented by linear, technology-specific marginal damage coefficients. The coefficients for renewable, hydropower, nuclear, coal, fuel oil, and gas generation are set to $0$, $0$, $0$, $90$, $95$, and $110$~\$/MWh, respectively. These stylized coefficients are selected so that, together with the underlying operating costs, the resulting social-cost merit order is consistent with the intended simulation design.

To create meaningful congestion, the thermal limits of all transmission lines are reduced to 50\% of their nominal values. The coefficients of the utility functions increase by 4\% per time step, while maximum demand increases by 2.5\% per time step. Annualized investment costs are based on technology cost data incorporating plant lifetime and total investment-cost components \citep{eia2022cost}. To isolate the strategic capacity effect, the baseline analysis assumes the same investment cost per MW across technologies. Unless otherwise stated, all monetary values are reported in dollars.

We first compare two spot-market models over the 20 time steps: (i) competitive market clearing that ignores pollution damages in the dispatch objective, and (ii) socially optimal market clearing that includes pollution damages, which is equivalent to competitive clearing under the proposed Pigouvian tax. For comparability, social welfare in both cases is evaluated using the same welfare measure, including pollution damages.

Figure~\ref{fig:optimal_competitive_market_price} shows the average nodal price\footnote{The average is calculated across all buses of the IEEE 24-bus system.} over time, while Figure~\ref{fig:optimal_competitive_market_welfare} reports the corresponding social welfare. Across the 20 time steps, the optimal dispatch produces a small but positive welfare improvement of approximately 0.10\% relative to competitive dispatch. The difference is modest because the system already contains a substantial share of low-emission generation, but it persists because the pollution-damage term affects the dispatch order when the system becomes more constrained. Figure~\ref{fig:optimal_competitive_market_tax} presents the tax payments of two sample producers. The levels and trends differ because the producers own different technology portfolios. Producers with more polluting generation face higher tax payments. As demand increases an additional polluting units enter the dispatch, their tax payments also generally increase.

\begin{figure*}[t]
    \centering
    \begin{subfigure}[b]{0.32\textwidth}
        \centering
\begin{tikzpicture}[scale=0.5]

\definecolor{darkgray176}{RGB}{176,176,176}
\definecolor{gray126155100}{RGB}{126,155,100}
\definecolor{lightgray204}{RGB}{204,204,204}
\definecolor{slategray129100155}{RGB}{129,100,155}

\begin{axis}[
height=6.5cm,
legend cell align={left},
legend style={
  fill opacity=0.8,
  draw opacity=1,
  text opacity=1,
  at={(0.03,0.97)},
  anchor=north west,
  draw=lightgray204
},
tick align=outside,
tick pos=left,
width=8.5cm,
x grid style={darkgray176},
xlabel={Time step (yr)},
xmin=0, xmax=21,
xtick style={color=black},
y grid style={darkgray176},
ylabel={Average price (\$/MWh)},
ymin=0, ymax=460,
ytick style={color=black}
]
\draw[draw=none,fill=gray126155100] (axis cs:0.65,0) rectangle (axis cs:0.95,82.4021012416428);
\addlegendimage{ybar,ybar legend,draw=none,fill=gray126155100}
\addlegendentry{Optimal}

\draw[draw=none,fill=gray126155100] (axis cs:1.65,0) rectangle (axis cs:1.95,110.139563862928);
\draw[draw=none,fill=gray126155100] (axis cs:2.65,0) rectangle (axis cs:2.95,114.545146417445);
\draw[draw=none,fill=gray126155100] (axis cs:3.65,0) rectangle (axis cs:3.95,115.447375886525);
\draw[draw=none,fill=gray126155100] (axis cs:4.65,0) rectangle (axis cs:4.95,107.606115369887);
\draw[draw=none,fill=gray126155100] (axis cs:5.65,0) rectangle (axis cs:5.95,124.912971683662);
\draw[draw=none,fill=gray126155100] (axis cs:6.65,0) rectangle (axis cs:6.95,129.909490551009);
\draw[draw=none,fill=gray126155100] (axis cs:7.65,0) rectangle (axis cs:7.95,139.27089242288);
\draw[draw=none,fill=gray126155100] (axis cs:8.65,0) rectangle (axis cs:8.95,122.385400165216);
\draw[draw=none,fill=gray126155100] (axis cs:9.65,0) rectangle (axis cs:9.95,141.672235010554);
\draw[draw=none,fill=gray126155100] (axis cs:10.65,0) rectangle (axis cs:10.95,150.141509854293);
\draw[draw=none,fill=gray126155100] (axis cs:11.65,0) rectangle (axis cs:11.95,169.533533095543);
\draw[draw=none,fill=gray126155100] (axis cs:12.65,0) rectangle (axis cs:12.95,201.301187140083);
\draw[draw=none,fill=gray126155100] (axis cs:13.65,0) rectangle (axis cs:13.95,245.527893722093);
\draw[draw=none,fill=gray126155100] (axis cs:14.65,0) rectangle (axis cs:14.95,267.214290395087);
\draw[draw=none,fill=gray126155100] (axis cs:15.65,0) rectangle (axis cs:15.95,281.643498923053);
\draw[draw=none,fill=gray126155100] (axis cs:16.65,0) rectangle (axis cs:16.95,304.3797647456);
\draw[draw=none,fill=gray126155100] (axis cs:17.65,0) rectangle (axis cs:17.95,360.110173224795);
\draw[draw=none,fill=gray126155100] (axis cs:18.65,0) rectangle (axis cs:18.95,410.374385152389);
\draw[draw=none,fill=gray126155100] (axis cs:19.65,0) rectangle (axis cs:19.95,428.264772103696);
\draw[draw=none,fill=slategray129100155] (axis cs:1.05,0) rectangle (axis cs:1.35,43.5745708967438);
\addlegendimage{ybar,ybar legend,draw=none,fill=slategray129100155}
\addlegendentry{Competetive}

\draw[draw=none,fill=slategray129100155] (axis cs:2.05,0) rectangle (axis cs:2.35,53.7380569867433);
\draw[draw=none,fill=slategray129100155] (axis cs:3.05,0) rectangle (axis cs:3.35,61.618963449861);
\draw[draw=none,fill=slategray129100155] (axis cs:4.05,0) rectangle (axis cs:4.35,66.5067341994934);
\draw[draw=none,fill=slategray129100155] (axis cs:5.05,0) rectangle (axis cs:5.35,78.7267264624359);
\draw[draw=none,fill=slategray129100155] (axis cs:6.05,0) rectangle (axis cs:6.35,83.2161403236631);
\draw[draw=none,fill=slategray129100155] (axis cs:7.05,0) rectangle (axis cs:7.35,36.5269144996448);
\draw[draw=none,fill=slategray129100155] (axis cs:8.05,0) rectangle (axis cs:8.35,107.782646593815);
\draw[draw=none,fill=slategray129100155] (axis cs:9.05,0) rectangle (axis cs:9.35,124.95228121063);
\draw[draw=none,fill=slategray129100155] (axis cs:10.05,0) rectangle (axis cs:10.35,132.223694585735);
\draw[draw=none,fill=slategray129100155] (axis cs:11.05,0) rectangle (axis cs:11.35,151.949936954489);
\draw[draw=none,fill=slategray129100155] (axis cs:12.05,0) rectangle (axis cs:12.35,166.399305524545);
\draw[draw=none,fill=slategray129100155] (axis cs:13.05,0) rectangle (axis cs:13.35,96.2516346428272);
\draw[draw=none,fill=slategray129100155] (axis cs:14.05,0) rectangle (axis cs:14.35,242.148456052117);
\draw[draw=none,fill=slategray129100155] (axis cs:15.05,0) rectangle (axis cs:15.35,261.245570205732);
\draw[draw=none,fill=slategray129100155] (axis cs:16.05,0) rectangle (axis cs:16.35,277.737744397423);
\draw[draw=none,fill=slategray129100155] (axis cs:17.05,0) rectangle (axis cs:17.35,315.230209181819);
\draw[draw=none,fill=slategray129100155] (axis cs:18.05,0) rectangle (axis cs:18.35,360.110173224795);
\draw[draw=none,fill=slategray129100155] (axis cs:19.05,0) rectangle (axis cs:19.35,406.953483882903);
\draw[draw=none,fill=slategray129100155] (axis cs:20.05,0) rectangle (axis cs:20.35,428.264772103696);
\end{axis}

\end{tikzpicture}
        \vspace{-0.5cm}
        \caption{Average market price}
        \label{fig:optimal_competitive_market_price}
    \end{subfigure}
    \hfill
    \begin{subfigure}[b]{0.31\textwidth}
        \centering
\begin{tikzpicture}[scale=0.5]

\definecolor{darkgray176}{RGB}{176,176,176}
\definecolor{gray126155100}{RGB}{126,155,100}
\definecolor{lightgray204}{RGB}{204,204,204}
\definecolor{slategray129100155}{RGB}{129,100,155}

\begin{axis}[
height=6.5cm,
legend cell align={left},
legend style={
  fill opacity=0.8,
  draw opacity=1,
  text opacity=1,
  at={(0.03,0.97)},
  anchor=north west,
  draw=lightgray204
},
tick align=outside,
tick pos=left,
width=8.5cm,
x grid style={darkgray176},
xlabel={Time step (yr)},
xmin=0, xmax=21,
xtick style={color=black},
y grid style={darkgray176},
ylabel={Social Welfare (M\$)},
ymin=0, ymax=8.5,
ytick style={color=black}
]
\draw[draw=none,fill=gray126155100] (axis cs:0.65,0) rectangle (axis cs:0.95,2.22448378825112);
\addlegendimage{ybar,ybar legend,draw=none,fill=gray126155100}
\addlegendentry{Optimal}

\draw[draw=none,fill=gray126155100] (axis cs:1.65,0) rectangle (axis cs:1.95,2.36995950354652);
\draw[draw=none,fill=gray126155100] (axis cs:2.65,0) rectangle (axis cs:2.95,2.52487377546179);
\draw[draw=none,fill=gray126155100] (axis cs:3.65,0) rectangle (axis cs:3.95,2.69071010871536);
\draw[draw=none,fill=gray126155100] (axis cs:4.65,0) rectangle (axis cs:4.95,2.86780931272036);
\draw[draw=none,fill=gray126155100] (axis cs:5.65,0) rectangle (axis cs:5.95,3.0569230371958);
\draw[draw=none,fill=gray126155100] (axis cs:6.65,0) rectangle (axis cs:6.95,3.25874093007122);
\draw[draw=none,fill=gray126155100] (axis cs:7.65,0) rectangle (axis cs:7.95,3.47361864677811);
\draw[draw=none,fill=gray126155100] (axis cs:8.65,0) rectangle (axis cs:8.95,3.70229891499233);
\draw[draw=none,fill=gray126155100] (axis cs:9.65,0) rectangle (axis cs:9.95,3.94603677379011);
\draw[draw=none,fill=gray126155100] (axis cs:10.65,0) rectangle (axis cs:10.95,4.20546745279768);
\draw[draw=none,fill=gray126155100] (axis cs:11.65,0) rectangle (axis cs:11.95,4.48024751830237);
\draw[draw=none,fill=gray126155100] (axis cs:12.65,0) rectangle (axis cs:12.95,4.77083166306594);
\draw[draw=none,fill=gray126155100] (axis cs:13.65,0) rectangle (axis cs:13.95,5.07745397011097);
\draw[draw=none,fill=gray126155100] (axis cs:14.65,0) rectangle (axis cs:14.95,5.40302988988856);
\draw[draw=none,fill=gray126155100] (axis cs:15.65,0) rectangle (axis cs:15.95,5.74841549181122);
\draw[draw=none,fill=gray126155100] (axis cs:16.65,0) rectangle (axis cs:16.95,6.11522524855884);
\draw[draw=none,fill=gray126155100] (axis cs:17.65,0) rectangle (axis cs:17.95,6.50245054235143);
\draw[draw=none,fill=gray126155100] (axis cs:18.65,0) rectangle (axis cs:18.95,6.91151197496707);
\draw[draw=none,fill=gray126155100] (axis cs:19.65,0) rectangle (axis cs:19.95,7.34311502583015);
\draw[draw=none,fill=slategray129100155] (axis cs:1.05,0) rectangle (axis cs:1.35,2.20966335160774);
\addlegendimage{ybar,ybar legend,draw=none,fill=slategray129100155}
\addlegendentry{Competetive}

\draw[draw=none,fill=slategray129100155] (axis cs:2.05,0) rectangle (axis cs:2.35,2.35323855393705);
\draw[draw=none,fill=slategray129100155] (axis cs:3.05,0) rectangle (axis cs:3.35,2.51197007518026);
\draw[draw=none,fill=slategray129100155] (axis cs:4.05,0) rectangle (axis cs:4.35,2.68060094313681);
\draw[draw=none,fill=slategray129100155] (axis cs:5.05,0) rectangle (axis cs:5.35,2.8603949813754);
\draw[draw=none,fill=slategray129100155] (axis cs:6.05,0) rectangle (axis cs:6.35,3.05199329144813);
\draw[draw=none,fill=slategray129100155] (axis cs:7.05,0) rectangle (axis cs:7.35,3.25575747826892);
\draw[draw=none,fill=slategray129100155] (axis cs:8.05,0) rectangle (axis cs:8.35,3.47176745722348);
\draw[draw=none,fill=slategray129100155] (axis cs:9.05,0) rectangle (axis cs:9.35,3.70095757664001);
\draw[draw=none,fill=slategray129100155] (axis cs:10.05,0) rectangle (axis cs:10.35,3.94494600137425);
\draw[draw=none,fill=slategray129100155] (axis cs:11.05,0) rectangle (axis cs:11.35,4.20440343555175);
\draw[draw=none,fill=slategray129100155] (axis cs:12.05,0) rectangle (axis cs:12.35,4.47920171156943);
\draw[draw=none,fill=slategray129100155] (axis cs:13.05,0) rectangle (axis cs:13.35,4.76980736038963);
\draw[draw=none,fill=slategray129100155] (axis cs:14.05,0) rectangle (axis cs:14.35,5.0764552398833);
\draw[draw=none,fill=slategray129100155] (axis cs:15.05,0) rectangle (axis cs:15.35,5.40205933119491);
\draw[draw=none,fill=slategray129100155] (axis cs:16.05,0) rectangle (axis cs:16.35,5.74745295130584);
\draw[draw=none,fill=slategray129100155] (axis cs:17.05,0) rectangle (axis cs:17.35,6.11433615517521);
\draw[draw=none,fill=slategray129100155] (axis cs:18.05,0) rectangle (axis cs:18.35,6.50161047069488);
\draw[draw=none,fill=slategray129100155] (axis cs:19.05,0) rectangle (axis cs:19.35,6.91072231064646);
\draw[draw=none,fill=slategray129100155] (axis cs:20.05,0) rectangle (axis cs:20.35,7.34238174856522);
\end{axis}

\end{tikzpicture}
        \vspace{-0.5cm}
        \caption{Social welfare}
        \label{fig:optimal_competitive_market_welfare}
    \end{subfigure}
    \hfill
    \begin{subfigure}[b]{0.32\textwidth}
        \centering
\begin{tikzpicture}[scale=0.5]

\definecolor{darkgray176}{RGB}{176,176,176}
\definecolor{gray126155100}{RGB}{126,155,100}
\definecolor{lightgray204}{RGB}{204,204,204}
\definecolor{slategray129100155}{RGB}{129,100,155}

\begin{axis}[
height=6.5cm,
legend cell align={left},
legend style={
  fill opacity=0.8,
  draw opacity=1,
  text opacity=1,
  at={(0.03,0.97)},
  anchor=north west,
  draw=lightgray204
},
tick align=outside,
tick pos=left,
width=8.2cm,
x grid style={darkgray176},
xlabel={Time step (yr)},
xmin=0, xmax=21,
xtick style={color=black},
y grid style={darkgray176},
ylabel={Proposed Tax (k\$)},
ymin=0, ymax=72,
ytick style={color=black}
]
\draw[draw=none,fill=gray126155100] (axis cs:0.65,0) rectangle (axis cs:0.95,16.649262470353);
\addlegendimage{ybar,ybar legend,draw=none,fill=gray126155100}
\addlegendentry{Producer 2}

\draw[draw=none,fill=gray126155100] (axis cs:1.65,0) rectangle (axis cs:1.95,17.1355592316536);
\draw[draw=none,fill=gray126155100] (axis cs:2.65,0) rectangle (axis cs:2.95,17.3470182374987);
\draw[draw=none,fill=gray126155100] (axis cs:3.65,0) rectangle (axis cs:3.95,17.5340093102854);
\draw[draw=none,fill=gray126155100] (axis cs:4.65,0) rectangle (axis cs:4.95,17.7651748299472);
\draw[draw=none,fill=gray126155100] (axis cs:5.65,0) rectangle (axis cs:5.95,18.0018651651033);
\draw[draw=none,fill=gray126155100] (axis cs:6.65,0) rectangle (axis cs:6.95,18.2430145202516);
\draw[draw=none,fill=gray126155100] (axis cs:7.65,0) rectangle (axis cs:7.95,18.8621812404667);
\draw[draw=none,fill=gray126155100] (axis cs:8.65,0) rectangle (axis cs:8.95,19.1213589069531);
\draw[draw=none,fill=gray126155100] (axis cs:9.65,0) rectangle (axis cs:9.95,20.7724297866735);
\draw[draw=none,fill=gray126155100] (axis cs:10.65,0) rectangle (axis cs:10.95,21.1262603885933);
\draw[draw=none,fill=gray126155100] (axis cs:11.65,0) rectangle (axis cs:11.95,21.4438045184334);
\draw[draw=none,fill=gray126155100] (axis cs:12.65,0) rectangle (axis cs:12.95,21.7661028811805);
\draw[draw=none,fill=gray126155100] (axis cs:13.65,0) rectangle (axis cs:13.95,22.086664171199);
\draw[draw=none,fill=gray126155100] (axis cs:14.65,0) rectangle (axis cs:14.95,22.4257607134438);
\draw[draw=none,fill=gray126155100] (axis cs:15.65,0) rectangle (axis cs:15.95,22.783138870897);
\draw[draw=none,fill=gray126155100] (axis cs:16.65,0) rectangle (axis cs:16.95,23.2005191374964);
\draw[draw=none,fill=gray126155100] (axis cs:17.65,0) rectangle (axis cs:17.95,23.5676076375061);
\draw[draw=none,fill=gray126155100] (axis cs:18.65,0) rectangle (axis cs:18.95,23.9334734750263);
\draw[draw=none,fill=gray126155100] (axis cs:19.65,0) rectangle (axis cs:19.95,24.3124555094334);
\draw[draw=none,fill=slategray129100155] (axis cs:1.05,0) rectangle (axis cs:1.35,18.3080961295876);
\addlegendimage{ybar,ybar legend,draw=none,fill=slategray129100155}
\addlegendentry{Producer 4}

\draw[draw=none,fill=slategray129100155] (axis cs:2.05,0) rectangle (axis cs:2.35,20.2607890220655);
\draw[draw=none,fill=slategray129100155] (axis cs:3.05,0) rectangle (axis cs:3.35,29.3325250210295);
\draw[draw=none,fill=slategray129100155] (axis cs:4.05,0) rectangle (axis cs:4.35,37.357154281818);
\draw[draw=none,fill=slategray129100155] (axis cs:5.05,0) rectangle (axis cs:5.35,42.9358166332677);
\draw[draw=none,fill=slategray129100155] (axis cs:6.05,0) rectangle (axis cs:6.35,47.6296718256052);
\draw[draw=none,fill=slategray129100155] (axis cs:7.05,0) rectangle (axis cs:7.35,52.0942568332884);
\draw[draw=none,fill=slategray129100155] (axis cs:8.05,0) rectangle (axis cs:8.35,53.3791530826918);
\draw[draw=none,fill=slategray129100155] (axis cs:9.05,0) rectangle (axis cs:9.35,54.6035961030725);
\draw[draw=none,fill=slategray129100155] (axis cs:10.05,0) rectangle (axis cs:10.35,55.9298100056294);
\draw[draw=none,fill=slategray129100155] (axis cs:11.05,0) rectangle (axis cs:11.35,56.145);
\draw[draw=none,fill=slategray129100155] (axis cs:12.05,0) rectangle (axis cs:12.35,56.145);
\draw[draw=none,fill=slategray129100155] (axis cs:13.05,0) rectangle (axis cs:13.35,56.145);
\draw[draw=none,fill=slategray129100155] (axis cs:14.05,0) rectangle (axis cs:14.35,56.145);
\draw[draw=none,fill=slategray129100155] (axis cs:15.05,0) rectangle (axis cs:15.35,56.145);
\draw[draw=none,fill=slategray129100155] (axis cs:16.05,0) rectangle (axis cs:16.35,56.145);
\draw[draw=none,fill=slategray129100155] (axis cs:17.05,0) rectangle (axis cs:17.35,56.145);
\draw[draw=none,fill=slategray129100155] (axis cs:18.05,0) rectangle (axis cs:18.35,56.145);
\draw[draw=none,fill=slategray129100155] (axis cs:19.05,0) rectangle (axis cs:19.35,56.145);
\draw[draw=none,fill=slategray129100155] (axis cs:20.05,0) rectangle (axis cs:20.35,56.145);
\end{axis}

\end{tikzpicture}
        \caption{Tax for two sample producers}
        \label{fig:optimal_competitive_market_tax}
    \end{subfigure}
    \vspace{-0.2cm}
    \caption{Comparison of optimal and competitive spot-market outcomes}
    \label{fig:optimal_competitive_market}
    \vspace{-0.2cm}
\end{figure*}

We next examine generation-capacity investment. We compare (i) a socially optimal investment model, in which the planner chooses capacity to maximize spot-market welfare net of investment costs, and (ii) a strategic investment model, in which producers choose capacity to maximize profit while anticipating the effect of capacity on future prices. The socially optimal investment outcome provides the benchmark that the complete tax--subsidy mechanism is designed to implement, whereas the strategic model represents the taxed market without the corrective investment subsidy. The corresponding linear formulations are presented in~\ref{sec:Appendix}. Figure~\ref{fig:optimal_strategic_investment_price} shows the average nodal price under the two investment models, and Figure~\ref{fig:optimal_strategic_investment_welfare} shows the corresponding social welfare. The strategic investment outcome produces a welfare loss of 4.25\% relative to the socially optimal benchmark. Total new generation-capacity investment is 65.77~MW in the optimal case and 1835.30~MW in the strategic case. The strategic outcome arises from individually optimal capacity decisions that account for the effect of capacity investment on market prices. Collectively, however, these decisions produce substantial and socially costly overinvestment.

Table~\ref{tab:investment-utilization} shows that the strategic case invests in all six technologies, whereas the optimal case adds only nuclear and fuel-oil capacity. Nuclear generation accounts for 1029.80~MW, or approximately 56\%, of total strategic investment. Including existing generation capacity, total installed capacity is 3470.77~MW in the optimal case and 5240.30~MW in the strategic case, representing an increase of approximately 51\%. However, the capacity-weighted utilization rate decreases from 84.84\% in the optimal case to 60.78\% in the strategic case. Aggregate utilized capacity increases by only 8.16\%, indicating that a large share of the additional strategically installed capacity remains underutilized.

\begin{figure*}[t]
    \centering
    \begin{subfigure}[b]{0.32\textwidth}
        \centering
\begin{tikzpicture}[scale=0.5]

\definecolor{darkgray176}{RGB}{176,176,176}
\definecolor{gray126155100}{RGB}{126,155,100}
\definecolor{lightgray204}{RGB}{204,204,204}
\definecolor{slategray129100155}{RGB}{129,100,155}

\begin{axis}[
height=6.5cm,
legend cell align={left},
legend style={
  fill opacity=0.8,
  draw opacity=1,
  text opacity=1,
  at={(0.03,0.97)},
  anchor=north west,
  draw=lightgray204
},
tick align=outside,
tick pos=left,
width=8.5cm,
x grid style={darkgray176},
xlabel={Time step (yr)},
xmin=0, xmax=21,
xtick style={color=black},
y grid style={darkgray176},
ylabel={Average price (\$/MWh)},
ymin=0, ymax=510,
ytick style={color=black}
]
\draw[draw=none,fill=gray126155100] (axis cs:0.65,0) rectangle (axis cs:0.95,99.208782051565);
\addlegendimage{ybar,ybar legend,draw=none,fill=gray126155100}
\addlegendentry{Optimal}

\draw[draw=none,fill=gray126155100] (axis cs:1.65,0) rectangle (axis cs:1.95,103.177133333628);
\draw[draw=none,fill=gray126155100] (axis cs:2.65,0) rectangle (axis cs:2.95,107.304218666973);
\draw[draw=none,fill=gray126155100] (axis cs:3.65,0) rectangle (axis cs:3.95,115.823806753482);
\draw[draw=none,fill=gray126155100] (axis cs:4.65,0) rectangle (axis cs:4.95,107.606115369887);
\draw[draw=none,fill=gray126155100] (axis cs:5.65,0) rectangle (axis cs:5.95,111.910359984682);
\draw[draw=none,fill=gray126155100] (axis cs:6.65,0) rectangle (axis cs:6.95,107.591169628286);
\draw[draw=none,fill=gray126155100] (axis cs:7.65,0) rectangle (axis cs:7.95,139.27089242288);
\draw[draw=none,fill=gray126155100] (axis cs:8.65,0) rectangle (axis cs:8.95,122.385400165216);
\draw[draw=none,fill=gray126155100] (axis cs:9.65,0) rectangle (axis cs:9.95,141.672235010554);
\draw[draw=none,fill=gray126155100] (axis cs:10.65,0) rectangle (axis cs:10.95,124.020363229573);
\draw[draw=none,fill=gray126155100] (axis cs:11.65,0) rectangle (axis cs:11.95,161.86326545448);
\draw[draw=none,fill=gray126155100] (axis cs:12.65,0) rectangle (axis cs:12.95,191.681719911384);
\draw[draw=none,fill=gray126155100] (axis cs:13.65,0) rectangle (axis cs:13.95,245.492096764675);
\draw[draw=none,fill=gray126155100] (axis cs:14.65,0) rectangle (axis cs:14.95,254.559148226642);
\draw[draw=none,fill=gray126155100] (axis cs:15.65,0) rectangle (axis cs:15.95,265.464663481328);
\draw[draw=none,fill=gray126155100] (axis cs:16.65,0) rectangle (axis cs:16.95,288.31890558407);
\draw[draw=none,fill=gray126155100] (axis cs:17.65,0) rectangle (axis cs:17.95,306.100260981508);
\draw[draw=none,fill=gray126155100] (axis cs:18.65,0) rectangle (axis cs:18.95,372.503349716943);
\draw[draw=none,fill=gray126155100] (axis cs:19.65,0) rectangle (axis cs:19.95,469.169311601144);
\draw[draw=none,fill=slategray129100155] (axis cs:1.05,0) rectangle (axis cs:1.35,62.2006342753518);
\addlegendimage{ybar,ybar legend,draw=none,fill=slategray129100155}
\addlegendentry{Strategic}

\draw[draw=none,fill=slategray129100155] (axis cs:2.05,0) rectangle (axis cs:2.35,72.4966276520697);
\draw[draw=none,fill=slategray129100155] (axis cs:3.05,0) rectangle (axis cs:3.35,77.4861191186283);
\draw[draw=none,fill=slategray129100155] (axis cs:4.05,0) rectangle (axis cs:4.35,82.8688457398043);
\draw[draw=none,fill=slategray129100155] (axis cs:5.05,0) rectangle (axis cs:5.35,79.56728100454);
\draw[draw=none,fill=slategray129100155] (axis cs:6.05,0) rectangle (axis cs:6.35,78.0549634102794);
\draw[draw=none,fill=slategray129100155] (axis cs:7.05,0) rectangle (axis cs:7.35,112.029464517158);
\draw[draw=none,fill=slategray129100155] (axis cs:8.05,0) rectangle (axis cs:8.35,105.721692149974);
\draw[draw=none,fill=slategray129100155] (axis cs:9.05,0) rectangle (axis cs:9.35,80.2762915785939);
\draw[draw=none,fill=slategray129100155] (axis cs:10.05,0) rectangle (axis cs:10.35,88.1138218617277);
\draw[draw=none,fill=slategray129100155] (axis cs:11.05,0) rectangle (axis cs:11.35,112.635388658166);
\draw[draw=none,fill=slategray129100155] (axis cs:12.05,0) rectangle (axis cs:12.35,96.8093049610713);
\draw[draw=none,fill=slategray129100155] (axis cs:13.05,0) rectangle (axis cs:13.35,137.540173268258);
\draw[draw=none,fill=slategray129100155] (axis cs:14.05,0) rectangle (axis cs:14.35,126.311019681038);
\draw[draw=none,fill=slategray129100155] (axis cs:15.05,0) rectangle (axis cs:15.35,143.088679837903);
\draw[draw=none,fill=slategray129100155] (axis cs:16.05,0) rectangle (axis cs:16.35,158.993096619108);
\draw[draw=none,fill=slategray129100155] (axis cs:17.05,0) rectangle (axis cs:17.35,203.169474239879);
\draw[draw=none,fill=slategray129100155] (axis cs:18.05,0) rectangle (axis cs:18.35,194.41118598338);
\draw[draw=none,fill=slategray129100155] (axis cs:19.05,0) rectangle (axis cs:19.35,218.058019657401);
\draw[draw=none,fill=slategray129100155] (axis cs:20.05,0) rectangle (axis cs:20.35,221.930412298106);
\end{axis}

\end{tikzpicture}
        \vspace{-0.5cm}
        \caption{Average market price}
        \label{fig:optimal_strategic_investment_price}
    \end{subfigure}
    \hfill
    \begin{subfigure}[b]{0.31\textwidth}
        \centering
\begin{tikzpicture}[scale=0.5]

\definecolor{darkgray176}{RGB}{176,176,176}
\definecolor{gray126155100}{RGB}{126,155,100}
\definecolor{lightgray204}{RGB}{204,204,204}
\definecolor{slategray129100155}{RGB}{129,100,155}

\begin{axis}[
height=6.5cm,
legend cell align={left},
legend style={
  fill opacity=0.8,
  draw opacity=1,
  text opacity=1,
  at={(0.03,0.97)},
  anchor=north west,
  draw=lightgray204
},
tick align=outside,
tick pos=left,
width=8.5cm,
x grid style={darkgray176},
xlabel={Time step (yr)},
xmin=0, xmax=21,
xtick style={color=black},
y grid style={darkgray176},
ylabel={Social Welfare (M\$)},
ymin=0, ymax=8.5,
ytick style={color=black}
]
\draw[draw=none,fill=gray126155100] (axis cs:0.65,0) rectangle (axis cs:0.95,2.21960825302131);
\addlegendimage{ybar,ybar legend,draw=none,fill=gray126155100}
\addlegendentry{Optimal}

\draw[draw=none,fill=gray126155100] (axis cs:1.65,0) rectangle (axis cs:1.95,2.36526663310802);
\draw[draw=none,fill=gray126155100] (axis cs:2.65,0) rectangle (axis cs:2.95,2.52031245528832);
\draw[draw=none,fill=gray126155100] (axis cs:3.65,0) rectangle (axis cs:3.95,2.68612809972271);
\draw[draw=none,fill=gray126155100] (axis cs:4.65,0) rectangle (axis cs:4.95,2.86339021115944);
\draw[draw=none,fill=gray126155100] (axis cs:5.65,0) rectangle (axis cs:5.95,3.05261544283828);
\draw[draw=none,fill=gray126155100] (axis cs:6.65,0) rectangle (axis cs:6.95,3.25460885696804);
\draw[draw=none,fill=gray126155100] (axis cs:7.65,0) rectangle (axis cs:7.95,3.46991846856712);
\draw[draw=none,fill=gray126155100] (axis cs:8.65,0) rectangle (axis cs:8.95,3.69859873678134);
\draw[draw=none,fill=gray126155100] (axis cs:9.65,0) rectangle (axis cs:9.95,3.94238218353794);
\draw[draw=none,fill=gray126155100] (axis cs:10.65,0) rectangle (axis cs:10.95,4.20228887179052);
\draw[draw=none,fill=gray126155100] (axis cs:11.65,0) rectangle (axis cs:11.95,4.47870094663089);
\draw[draw=none,fill=gray126155100] (axis cs:12.65,0) rectangle (axis cs:12.95,4.7712070282432);
\draw[draw=none,fill=gray126155100] (axis cs:13.65,0) rectangle (axis cs:13.95,5.08083410913059);
\draw[draw=none,fill=gray126155100] (axis cs:14.65,0) rectangle (axis cs:14.95,5.40732781967747);
\draw[draw=none,fill=gray126155100] (axis cs:15.65,0) rectangle (axis cs:15.95,5.75428006253977);
\draw[draw=none,fill=gray126155100] (axis cs:16.65,0) rectangle (axis cs:16.95,6.12308997692843);
\draw[draw=none,fill=gray126155100] (axis cs:17.65,0) rectangle (axis cs:17.95,6.51486964864262);
\draw[draw=none,fill=gray126155100] (axis cs:18.65,0) rectangle (axis cs:18.95,6.92911142591552);
\draw[draw=none,fill=gray126155100] (axis cs:19.65,0) rectangle (axis cs:19.95,7.36406334123372);
\draw[draw=none,fill=slategray129100155] (axis cs:1.05,0) rectangle (axis cs:1.35,1.99886098887677);
\addlegendimage{ybar,ybar legend,draw=none,fill=slategray129100155}
\addlegendentry{Strategic}

\draw[draw=none,fill=slategray129100155] (axis cs:2.05,0) rectangle (axis cs:2.35,2.14605722812236);
\draw[draw=none,fill=slategray129100155] (axis cs:3.05,0) rectangle (axis cs:3.35,2.30252380621404);
\draw[draw=none,fill=slategray129100155] (axis cs:4.05,0) rectangle (axis cs:4.35,2.46954432386684);
\draw[draw=none,fill=slategray129100155] (axis cs:5.05,0) rectangle (axis cs:5.35,2.64784714426275);
\draw[draw=none,fill=slategray129100155] (axis cs:6.05,0) rectangle (axis cs:6.35,2.83824640147943);
\draw[draw=none,fill=slategray129100155] (axis cs:7.05,0) rectangle (axis cs:7.35,3.04163789666768);
\draw[draw=none,fill=slategray129100155] (axis cs:8.05,0) rectangle (axis cs:8.35,3.25896922501103);
\draw[draw=none,fill=slategray129100155] (axis cs:9.05,0) rectangle (axis cs:9.35,3.49092424683156);
\draw[draw=none,fill=slategray129100155] (axis cs:10.05,0) rectangle (axis cs:10.35,3.73853859812892);
\draw[draw=none,fill=slategray129100155] (axis cs:11.05,0) rectangle (axis cs:11.35,4.00258107854044);
\draw[draw=none,fill=slategray129100155] (axis cs:12.05,0) rectangle (axis cs:12.35,4.28349394740879);
\draw[draw=none,fill=slategray129100155] (axis cs:13.05,0) rectangle (axis cs:13.35,4.58282428547331);
\draw[draw=none,fill=slategray129100155] (axis cs:14.05,0) rectangle (axis cs:14.35,4.90175607974031);
\draw[draw=none,fill=slategray129100155] (axis cs:15.05,0) rectangle (axis cs:15.35,5.24104976608265);
\draw[draw=none,fill=slategray129100155] (axis cs:16.05,0) rectangle (axis cs:16.35,5.60166475109092);
\draw[draw=none,fill=slategray129100155] (axis cs:17.05,0) rectangle (axis cs:17.35,5.9839351095115);
\draw[draw=none,fill=slategray129100155] (axis cs:18.05,0) rectangle (axis cs:18.35,6.38867276199242);
\draw[draw=none,fill=slategray129100155] (axis cs:19.05,0) rectangle (axis cs:19.35,6.81893151798516);
\draw[draw=none,fill=slategray129100155] (axis cs:20.05,0) rectangle (axis cs:20.35,7.27582290674395);
\end{axis}

\end{tikzpicture}
        \vspace{-0.5cm}
        \caption{Social welfare}
        \label{fig:optimal_strategic_investment_welfare}
    \end{subfigure}
    \hfill
    \begin{subfigure}[b]{0.32\textwidth}
        \centering
\begin{tikzpicture}[scale=0.5]

\definecolor{darkgray176}{RGB}{176,176,176}
\definecolor{gray126155100}{RGB}{126,155,100}
\definecolor{lightgray204}{RGB}{204,204,204}
\definecolor{slategray129100155}{RGB}{129,100,155}

\begin{axis}[
height=6.5cm,
legend cell align={left},
legend style={
  fill opacity=0.8,
  draw opacity=1,
  text opacity=1,
  at={(0.03,0.97)},
  anchor=north west,
  draw=lightgray204
},
tick align=outside,
tick pos=left,
width=8.5cm,
x grid style={darkgray176},
xlabel={Time step (yr)},
xmin=0, xmax=21,
xtick style={color=black},
y grid style={darkgray176},
ylabel={Profit (k\$)},
ymin=0, ymax=150,
ytick style={color=black}
]
\draw[draw=none,fill=gray126155100] (axis cs:0.65,0) rectangle (axis cs:0.95,29.885464064905);
\addlegendimage{ybar,ybar legend,draw=none,fill=gray126155100}
\addlegendentry{Optimal}

\draw[draw=none,fill=gray126155100] (axis cs:1.65,0) rectangle (axis cs:1.95,32.962403096772);
\draw[draw=none,fill=gray126155100] (axis cs:2.65,0) rectangle (axis cs:2.95,35.6083184841694);
\draw[draw=none,fill=gray126155100] (axis cs:3.65,0) rectangle (axis cs:3.95,40.0302572386409);
\draw[draw=none,fill=gray126155100] (axis cs:4.65,0) rectangle (axis cs:4.95,44.4873910220317);
\draw[draw=none,fill=gray126155100] (axis cs:5.65,0) rectangle (axis cs:5.95,43.9072740319637);
\draw[draw=none,fill=gray126155100] (axis cs:6.65,0) rectangle (axis cs:6.95,51.9510443609281);
\draw[draw=none,fill=gray126155100] (axis cs:7.65,0) rectangle (axis cs:7.95,49.8911952222911);
\draw[draw=none,fill=gray126155100] (axis cs:8.65,0) rectangle (axis cs:8.95,51.3984818240978);
\draw[draw=none,fill=gray126155100] (axis cs:9.65,0) rectangle (axis cs:9.95,54.9359308023944);
\draw[draw=none,fill=gray126155100] (axis cs:10.65,0) rectangle (axis cs:10.95,51.8426251760182);
\draw[draw=none,fill=gray126155100] (axis cs:11.65,0) rectangle (axis cs:11.95,53.7778803285188);
\draw[draw=none,fill=gray126155100] (axis cs:12.65,0) rectangle (axis cs:12.95,52.7714152308333);
\draw[draw=none,fill=gray126155100] (axis cs:13.65,0) rectangle (axis cs:13.95,57.2641162235913);
\draw[draw=none,fill=gray126155100] (axis cs:14.65,0) rectangle (axis cs:14.95,63.7398042844692);
\draw[draw=none,fill=gray126155100] (axis cs:15.65,0) rectangle (axis cs:15.95,74.1208678572035);
\draw[draw=none,fill=gray126155100] (axis cs:16.65,0) rectangle (axis cs:16.95,84.0795636583343);
\draw[draw=none,fill=gray126155100] (axis cs:17.65,0) rectangle (axis cs:17.95,103.47268838564);
\draw[draw=none,fill=gray126155100] (axis cs:18.65,0) rectangle (axis cs:18.95,114.995832427983);
\draw[draw=none,fill=gray126155100] (axis cs:19.65,0) rectangle (axis cs:19.95,130.282888164392);
\draw[draw=none,fill=slategray129100155] (axis cs:1.05,0) rectangle (axis cs:1.35,21.600797162037);
\addlegendimage{ybar,ybar legend,draw=none,fill=slategray129100155}
\addlegendentry{Strategic}

\draw[draw=none,fill=slategray129100155] (axis cs:2.05,0) rectangle (axis cs:2.35,27.9146491542457);
\draw[draw=none,fill=slategray129100155] (axis cs:3.05,0) rectangle (axis cs:3.35,21.0550131554765);
\draw[draw=none,fill=slategray129100155] (axis cs:4.05,0) rectangle (axis cs:4.35,24.9815687163852);
\draw[draw=none,fill=slategray129100155] (axis cs:5.05,0) rectangle (axis cs:5.35,32.9921507597297);
\draw[draw=none,fill=slategray129100155] (axis cs:6.05,0) rectangle (axis cs:6.35,28.5848568631571);
\draw[draw=none,fill=slategray129100155] (axis cs:7.05,0) rectangle (axis cs:7.35,23.89275654469);
\draw[draw=none,fill=slategray129100155] (axis cs:8.05,0) rectangle (axis cs:8.35,30.447409046079);
\draw[draw=none,fill=slategray129100155] (axis cs:9.05,0) rectangle (axis cs:9.35,24.1458874623801);
\draw[draw=none,fill=slategray129100155] (axis cs:10.05,0) rectangle (axis cs:10.35,14.4821538157413);
\draw[draw=none,fill=slategray129100155] (axis cs:11.05,0) rectangle (axis cs:11.35,21.6345185076958);
\draw[draw=none,fill=slategray129100155] (axis cs:12.05,0) rectangle (axis cs:12.35,37.7236355752516);
\draw[draw=none,fill=slategray129100155] (axis cs:13.05,0) rectangle (axis cs:13.35,28.2442573239987);
\draw[draw=none,fill=slategray129100155] (axis cs:14.05,0) rectangle (axis cs:14.35,17.5834895927824);
\draw[draw=none,fill=slategray129100155] (axis cs:15.05,0) rectangle (axis cs:15.35,14.9477141975258);
\draw[draw=none,fill=slategray129100155] (axis cs:16.05,0) rectangle (axis cs:16.35,32.5419047802538);
\draw[draw=none,fill=slategray129100155] (axis cs:17.05,0) rectangle (axis cs:17.35,34.1207056363742);
\draw[draw=none,fill=slategray129100155] (axis cs:18.05,0) rectangle (axis cs:18.35,14.1611033792201);
\draw[draw=none,fill=slategray129100155] (axis cs:19.05,0) rectangle (axis cs:19.35,8.56878263411925);
\draw[draw=none,fill=slategray129100155] (axis cs:20.05,0) rectangle (axis cs:20.35,14.9969720637612);
\end{axis}

\end{tikzpicture}
        \vspace{-0.5cm}
        \caption{Profit of a sample producer}
        \label{fig:optimal_strategic_investment_profit}
    \end{subfigure}
    \vspace{-0.2cm}
    \caption{Comparison of optimal and strategic investments}
    \label{fig:optimal_strategic_investment}
    \vspace{-0.4cm}
\end{figure*}

\begin{table}[t]
    \centering
    \caption{Investment and utilization by generation technology.}
    \label{tab:investment-utilization}

    \scriptsize
    \setlength{\tabcolsep}{4pt}
    \renewcommand{\arraystretch}{1.15}

    \begin{tabular}{lrrrr}
        \toprule
        \textbf{Technology}
        & \multicolumn{2}{c}{\textbf{New investment (MW)}}
        & \multicolumn{2}{c}{\textbf{Utilization (\%)}} \\
        \cmidrule(lr){2-3}
        \cmidrule(lr){4-5}
        & \textbf{Optimal}
        & \textbf{Strategic}
        & \textbf{Optimal}
        & \textbf{Strategic} \\
        \midrule
        Renewable  & 0.00  & 130.00  & 100.00 & 97.91 \\
        Hydropower & 0.00  & 200.00  & 62.24  & 16.83 \\
        Nuclear    & 39.11 & 1029.80 & 100.00 & 85.35 \\
        Coal       & 0.00  & 120.00  & 72.36  & 51.69 \\
        Fuel oil   & 26.66 & 295.50  & 83.47  & 41.07 \\
        Gas        & 0.00  & 60.00   & 72.23  & 24.18 \\
        \midrule
        \textbf{Total investment}
        & \textbf{65.77}
        & \textbf{1835.30}
        & -- & -- \\
        \bottomrule
    \end{tabular}
\end{table}

\begin{table}[t]
    \centering
    \caption{Decomposition of social welfare under optimal and strategic
    investment (\$).}
    \label{tab:welfare-decomposition}

    \scriptsize
    \setlength{\tabcolsep}{6pt}
    \renewcommand{\arraystretch}{1.15}

    \begin{tabular}{lrrr}
        \toprule
        \textbf{Welfare component}
        & \textbf{Optimal}
        & \textbf{Strategic}
        & \makecell{\textbf{Optimal minus}\\\textbf{strategic}} \\
        \midrule
        Total utility
        & 89,574,469.63
        & 90,433,450.05
        & -858,980.43 \\
        Operating cost
        & 1,313,032.06
        & 1,172,311.45
        & 140,720.61 \\
        Pollution damage
        & 1,390,500.33
        & 1,092,231.32
        & 298,269.01 \\
        Investment cost
        & 172,334.67
        & 5,155,025.22
        & -4,982,690.55 \\
        \midrule
        \textbf{Total social welfare}
        & \textbf{86,698,602.57}
        & \textbf{83,013,882.06}
        & \textbf{3,684,720.51} \\
        \bottomrule
    \end{tabular}
\end{table}

Table~\ref{tab:welfare-decomposition} shows that the strategic case produces approximately \$0.86 million more in total utility and incurs approximately \$0.14 million less in operating costs and \$0.30 million less in pollution damage. However, its investment cost is approximately \$4.98 million higher. This difference dominates the other welfare components, resulting in a welfare loss of approximately \$3.68 million, or 4.25\% relative to the optimal benchmark. Thus, the additional strategic capacity provides some operating and environmental benefits, but these benefits are insufficient to offset the cost of substantial underutilized investment. Figure~\ref{fig:optimal_strategic_investment_profit} presents the profit of a sample producer over time. In the optimal investment case, the producer's profit generally increases as demand grows. In the strategic case, profit does not follow the same pattern because strategic investment changes both dispatch and prices. Over the complete horizon, aggregate producer profit in the optimal investment case is 116.13\% higher than in the strategic investment case, measured relative to strategic profit. This result illustrates a collective overinvestment effect: although each strategic investment decision is individually optimal given the decisions of other producers, the resulting equilibrium can reduce both total welfare and aggregate producer profit.

We also examine the budget balance of the proposed mechanism in the baseline case. Figure~\ref{fig:budget-balance} compares annual pollution-tax revenues and investment-subsidy payments. Under the normalization used in the numerical study, no additional constant participation transfers are included in this calculation. Tax revenues exceed subsidy payments in several years, but the mechanism is not budget-balanced over the complete 20-year horizon. Cumulative tax revenues are $3815.03$~k\$, compared with cumulative subsidy payments of $4963.67$~k\$. Tax revenues therefore cover 76.86\% of the subsidy payments, leaving a net deficit of $1148.64$~k\$. An additional financing or cost-recovery rule is therefore required in the baseline case.

\begin{figure}[t]
    \centering
    \begin{tikzpicture}[scale=0.7]

\definecolor{darkgray176}{RGB}{176,176,176}
\definecolor{gray126155100}{RGB}{126,155,100}
\definecolor{lightgray204}{RGB}{204,204,204}
\definecolor{slategray129100155}{RGB}{129,100,155}

\begin{axis}[
height=6.5cm,
legend cell align={left},
legend style={
  fill opacity=0.8,
  draw opacity=1,
  text opacity=1,
  at={(0.03,0.97)},
  anchor=north west,
  draw=lightgray204
},
tick align=outside,
tick pos=left,
width=8.6cm,
x grid style={darkgray176},
xlabel={Time step (yr)},
xmin=0, xmax=21,
xtick={1,2,4,6,8,10,12,14,16,18,20},
xtick style={color=black},
y grid style={darkgray176},
ylabel={Payment (k\$)},
ymin=0, ymax=600,
ytick style={color=black}
]

\draw[draw=none,fill=gray126155100] (axis cs:0.65,0) rectangle (axis cs:0.95,81.54);
\addlegendimage{ybar,ybar legend,draw=none,fill=gray126155100}
\addlegendentry{Total levied tax}

\draw[draw=none,fill=gray126155100] (axis cs:1.65,0) rectangle (axis cs:1.95,89.89);
\draw[draw=none,fill=gray126155100] (axis cs:2.65,0) rectangle (axis cs:2.95,101.17);
\draw[draw=none,fill=gray126155100] (axis cs:3.65,0) rectangle (axis cs:3.95,144.99);
\draw[draw=none,fill=gray126155100] (axis cs:4.65,0) rectangle (axis cs:4.95,160.47);
\draw[draw=none,fill=gray126155100] (axis cs:5.65,0) rectangle (axis cs:5.95,174.01);
\draw[draw=none,fill=gray126155100] (axis cs:6.65,0) rectangle (axis cs:6.95,192.79);
\draw[draw=none,fill=gray126155100] (axis cs:7.65,0) rectangle (axis cs:7.95,199.24);
\draw[draw=none,fill=gray126155100] (axis cs:8.65,0) rectangle (axis cs:8.95,203.70);
\draw[draw=none,fill=gray126155100] (axis cs:9.65,0) rectangle (axis cs:9.95,212.71);
\draw[draw=none,fill=gray126155100] (axis cs:10.65,0) rectangle (axis cs:10.95,222.27);
\draw[draw=none,fill=gray126155100] (axis cs:11.65,0) rectangle (axis cs:11.95,223.03);
\draw[draw=none,fill=gray126155100] (axis cs:12.65,0) rectangle (axis cs:12.95,223.68);
\draw[draw=none,fill=gray126155100] (axis cs:13.65,0) rectangle (axis cs:13.95,224.34);
\draw[draw=none,fill=gray126155100] (axis cs:14.65,0) rectangle (axis cs:14.95,225.00);
\draw[draw=none,fill=gray126155100] (axis cs:15.65,0) rectangle (axis cs:15.95,225.69);
\draw[draw=none,fill=gray126155100] (axis cs:16.65,0) rectangle (axis cs:16.95,226.44);
\draw[draw=none,fill=gray126155100] (axis cs:17.65,0) rectangle (axis cs:17.95,227.24);
\draw[draw=none,fill=gray126155100] (axis cs:18.65,0) rectangle (axis cs:18.95,228.02);
\draw[draw=none,fill=gray126155100] (axis cs:19.65,0) rectangle (axis cs:19.95,228.79);

\draw[draw=none,fill=slategray129100155] (axis cs:1.05,0) rectangle (axis cs:1.35,108.24);
\addlegendimage{ybar,ybar legend,draw=none,fill=slategray129100155}
\addlegendentry{Total subsidy}

\draw[draw=none,fill=slategray129100155] (axis cs:2.05,0) rectangle (axis cs:2.35,112.14);
\draw[draw=none,fill=slategray129100155] (axis cs:3.05,0) rectangle (axis cs:3.35,119.94);
\draw[draw=none,fill=slategray129100155] (axis cs:4.05,0) rectangle (axis cs:4.35,70.19);
\draw[draw=none,fill=slategray129100155] (axis cs:5.05,0) rectangle (axis cs:5.35,93.31);
\draw[draw=none,fill=slategray129100155] (axis cs:6.05,0) rectangle (axis cs:6.35,100.76);
\draw[draw=none,fill=slategray129100155] (axis cs:7.05,0) rectangle (axis cs:7.35,123.25);
\draw[draw=none,fill=slategray129100155] (axis cs:8.05,0) rectangle (axis cs:8.35,94.28);
\draw[draw=none,fill=slategray129100155] (axis cs:9.05,0) rectangle (axis cs:9.35,198.99);
\draw[draw=none,fill=slategray129100155] (axis cs:10.05,0) rectangle (axis cs:10.35,178.09);
\draw[draw=none,fill=slategray129100155] (axis cs:11.05,0) rectangle (axis cs:11.35,300.01);
\draw[draw=none,fill=slategray129100155] (axis cs:12.05,0) rectangle (axis cs:12.35,246.63);
\draw[draw=none,fill=slategray129100155] (axis cs:13.05,0) rectangle (axis cs:13.35,323.09);
\draw[draw=none,fill=slategray129100155] (axis cs:14.05,0) rectangle (axis cs:14.35,276.37);
\draw[draw=none,fill=slategray129100155] (axis cs:15.05,0) rectangle (axis cs:15.35,322.50);
\draw[draw=none,fill=slategray129100155] (axis cs:16.05,0) rectangle (axis cs:16.35,387.60);
\draw[draw=none,fill=slategray129100155] (axis cs:17.05,0) rectangle (axis cs:17.35,415.91);
\draw[draw=none,fill=slategray129100155] (axis cs:18.05,0) rectangle (axis cs:18.35,547.29);
\draw[draw=none,fill=slategray129100155] (axis cs:19.05,0) rectangle (axis cs:19.35,502.84);
\draw[draw=none,fill=slategray129100155] (axis cs:20.05,0) rectangle (axis cs:20.35,442.24);

\end{axis}
\end{tikzpicture}
    \caption{Yearly tax revenues and subsidy payments in the baseline case.}
    \label{fig:budget-balance}
\end{figure}

We examine the sensitivity of the investment results to three central inputs: generation-investment costs, the upper bounds on new investment, and the load level. The corresponding multipliers are denoted by $\alpha_I$, $\alpha_K$, and $\alpha_D$, respectively. In each experiment, one input is varied while all remaining data are held at their baseline values. The strategic welfare loss is calculated relative to the corresponding socially optimal outcome as $100\,\frac{W^{\mathrm{opt}}-W^{\mathrm{str}}}{W^{\mathrm{opt}}}$.

\begin{table}[t]
    \centering
    \caption{Sensitivity of investment and welfare outcomes.}
    \label{tab:investment-sensitivity}

    \scriptsize
    \setlength{\tabcolsep}{7pt}
    \renewcommand{\arraystretch}{1.15}

    \begin{tabular}{lrrr}
        \toprule
        \textbf{Multiplier}
        & \makecell{\textbf{Optimal}\\\textbf{investment (MW)}}
        & \makecell{\textbf{Strategic}\\\textbf{investment (MW)}}
        & \makecell{\textbf{Strategic}\\\textbf{welfare loss (\%)}} \\
        \midrule
        \multicolumn{4}{l}{
            \textbf{Investment-cost multiplier, $\alpha_I$}
        } \\
        $0.75$ & 201.73 & 1852.73 & 2.89 \\
        $1.00$ & 65.77  & 1835.30 & 4.25 \\
        $1.25$ & 11.61  & 1835.30 & 5.71 \\
        $1.50$ & 3.84   & 1835.30 & 7.20 \\

        \addlinespace[0.35em]
        \multicolumn{4}{l}{
            \textbf{Investment-bound multiplier, $\alpha_K$}
        } \\
        $0.75$ & 65.77 & 1558.67 & 3.44 \\
        $1.00$ & 65.77 & 1835.30 & 4.25 \\
        $1.25$ & 65.77 & 2046.44 & 4.90 \\
        $1.50$ & 65.77 & 2234.70 & 5.46 \\

        \addlinespace[0.35em]
        \multicolumn{4}{l}{
            \textbf{Load multiplier, $\alpha_D$}
        } \\
        $0.90$ & 0.00   & 1689.61 & 4.79 \\
        $1.00$ & 65.77  & 1835.30 & 4.25 \\
        $1.10$ & 202.69 & 1899.52 & 3.59 \\
        \bottomrule
    \end{tabular}
\end{table}

Increasing investment costs substantially reduces socially optimal investment, from 201.73~MW at $\alpha_I=0.75$ to 3.84~MW at $\alpha_I=1.50$. Strategic investment remains above 1835~MW throughout this range, and the associated welfare loss increases from 2.89\% to 7.20\%. Optimal investment is unaffected by the tested investment bounds, whereas strategic investment increases from 1558.67~MW at $\alpha_K=0.75$ to 2234.70~MW at $\alpha_K=1.50$. The magnitude of strategic overinvestment is therefore sensitive to the admissible investment limits. Higher load increases the social value of new capacity. Optimal investment rises from zero at $\alpha_D=0.90$ to 202.69~MW at $\alpha_D=1.10$, while strategic investment remains substantially higher. The strategic welfare loss decreases from 4.79\% to 3.59\% because a larger share of the additional capacity is utilized at higher load.

The potential welfare benefit of the proposed mechanism corresponds to the distortion between the strategic-investment equilibrium and the social-welfare benchmark. In the baseline case, this gap is approximately \$3.68 million, or 4.25\% of optimal welfare, and is driven primarily by the cost of strategically installed and underutilized capacity. The sensitivity results indicate that the gap becomes larger when investment is more costly, investment limits are less restrictive, or system loading is insufficient to utilize the additional capacity. Accordingly, the mechanism is most valuable in systems where capacity decisions materially affect market prices and produce substantial, costly, and weakly utilized strategic investment.

The sensitivity results show that the quantitative magnitude of strategic overinvestment depends on the calibration. Nevertheless, strategic investment exceeds socially optimal investment and produces lower welfare in every tested scenario. The numerical values should therefore be interpreted as scenario-dependent illustrations of a persistent strategic-investment distortion rather than as universal capacity predictions. We additionally examine the sensitivity of the mechanism to imperfect operational and emissions information. Let $e_g$ and $c_g$ denote the true emissions coefficient and operating-cost bid of generator $g$, respectively. We use perturbed values $\widehat e_g=\alpha_e e_g$ and $\widehat c_g=\alpha_b c_g$ in market-clearing and policy calculations. For the capacity-information experiment, define available capacity as $K_{g,t}^{\mathrm{av}} =A_{g,t}^{G}\widebar{P}_{g}^{G}$ and let the reported value be $\widehat K_{g,t}^{\mathrm{av}} =\alpha_A K_{g,t}^{\mathrm{av}}$. Ex post operating costs, pollution damages, and social welfare are evaluated using the true parameter values. We consider $\alpha_e\in\{0.90,0.95,1.00,1.05,1.10\}$, $\alpha_b\in\{1.00,1.05,1.10,1.20\}$, and $\alpha_A\in\{1.00,0.95,0.90,0.80\}$. For these experiments, welfare loss is measured relative to the full-information mechanism outcome as $100\,\frac{W^{\mathrm{full}}-W^{\mathrm{pert}}}{W^{\mathrm{full}}}$.

\begin{table}[t]
    \centering
    \caption{Sensitivity to imperfect emissions and operational information.}
    \label{tab:information-robustness}

    \scriptsize
    \setlength{\tabcolsep}{7pt}
    \renewcommand{\arraystretch}{1.15}

    \begin{tabular}{lrrr}
        \toprule
        \textbf{Multiplier}
        & \makecell{\textbf{Resulting}\\\textbf{investment (MW)}}
        & \makecell{\textbf{Pollution damage}\\\textbf{(million \$)}}
        & \makecell{\textbf{Welfare loss}\\\textbf{(\%)}} \\
        \midrule
        \multicolumn{4}{l}{
            \textbf{Emissions-coefficient multiplier, $\alpha_e$}
        } \\
        $0.90$ & 68.29 & 1.427 & 0.0022 \\
        $0.95$ & 68.82 & 1.410 & 0.0006 \\
        $1.00$ & 65.77 & 1.391 & 0.0000 \\
        $1.05$ & 65.77 & 1.375 & 0.0005 \\
        $1.10$ & 65.04 & 1.356 & 0.0022 \\

        \addlinespace[0.35em]
        \multicolumn{4}{l}{
            \textbf{Operating-cost bid multiplier, $\alpha_b$}
        } \\
        $1.00$ & 65.77 & 1.391 & 0.0000 \\
        $1.05$ & 65.77 & 1.382 & 0.0003 \\
        $1.10$ & 65.98 & 1.374 & 0.0010 \\
        $1.20$ & 66.47 & 1.354 & 0.0042 \\

        \addlinespace[0.35em]
        \multicolumn{4}{l}{
            \textbf{Available-capacity multiplier, $\alpha_A$}
        } \\
        $1.00$ & 65.77  & 1.391 & 0.0000 \\
        $0.95$ & 149.03 & 1.432 & 0.3208 \\
        $0.90$ & 237.45 & 1.477 & 0.6678 \\
        $0.80$ & 445.77 & 1.567 & 1.4555 \\
        \bottomrule
    \end{tabular}
\end{table}

Moderate errors in the emissions coefficients have limited effects on investment and welfare in the test system. Across the $\pm10\%$ range, total investment remains between 65.04 and 68.82~MW, and the welfare loss does not exceed 0.0023\%. Uniform increases in operating-cost bids also produce small welfare effects, with a maximum loss of 0.0042\% for a 20\% increase. Because a common bid multiplier largely preserves the relative merit order, this experiment should be interpreted as a reduced-form information stress test rather than as a complete model of strategic bidding. The results are considerably more sensitive to inaccurate available-capacity information. Reducing reported available capacity by 5\%, 10\%, and 20\% increases total investment to 149.03, 237.45, and 445.77~MW, respectively. The corresponding pollution damage increases by 3.0\%, 6.2\%, and 12.7\%, while the welfare loss rises to 0.3208\%, 0.6678\%, and 1.4555\%. Accurate and verifiable capacity-availability information is therefore particularly important for implementing the proposed mechanism.
\vspace{-0.3cm}
\section{Conclusion}
\label{sec:conclusion}

This paper studied two related inefficiencies in wholesale electricity markets. First, competitive spot-market clearing may ignore pollution damages and lead to inefficient dispatch. We derived a generator-specific Pigouvian tax that charges each generator for its marginal contribution to pollution damages and, under the stated assumptions, aligns profit-maximizing dispatch with the socially optimal dispatch while respecting operational and network constraints. Second, because installed generation capacity affects future dispatch and spot prices, producers may invest strategically and deviate from the socially optimal capacity level. We therefore derived a generator-specific subsidy based on the producer's contribution to consumer surplus, calculated through counterfactual market re-clearing, which corrects the effect of capacity investment on market prices and aligns private investment incentives with social welfare. The analytical and numerical results illustrate both instruments and show that strategic overinvestment can reduce welfare through costly and underutilized capacity. The mechanism does not require new reports of complete private cost functions beyond existing market submissions, but it requires verified emissions and technical data, regulator-specified pollution-damage functions, counterfactual re-clearing, and a credible financing arrangement. Future research may jointly model operational and investment market power, assess the proposed mechanism in markets with energy-storage operation, develop budget-balanced financing rules, and design scalable methods for counterfactual market re-clearing in real-life power systems. Energy storage can alter dispatch, congestion, and market prices and therefore may also affect generation-investment incentives and the resulting tax and subsidy values \citep{khastieva2019value}. Extending the framework to include storage charging and discharging decisions is therefore an important direction for future work.

\appendix
\vspace{-0.3cm}
\section{Linear Model}
\label{sec:Appendix}
This appendix presents the linear formulation used in the numerical experiments. Generator operating costs, demand utilities, and pollution damages are represented by piecewise-linear functions. For notational compactness, each piecewise-linear segment is treated as a separate generation or demand block with a constant marginal coefficient.

We consider sets of generators $\mathcal{G}$, demands $\mathcal{D}$, transmission lines $\mathcal{L}$, and time steps $\mathcal{T}$. Let $P_{g,t}^{G}$ and $P_{d,t}^{D}$ denote generator output and demand consumption in time step $t$. Let $b_{g,t}^{G}$ be the marginal operating-cost coefficient of generator $g$, and let $b_{d,t}^{D}$ be the marginal utility coefficient of demand $d$. Existing generation capacity is denoted by $\widebar{P}_{g}^{G0}$, and $\Delta\widebar{P}_{g}^{G}$ denotes additional capacity investment. Transmission constraints are represented using the PTDF coefficients $H_{g,l}^{G}$ and $H_{d,l}^{D}$.

\vspace{-0.2cm}
\subsection{Competitive and socially optimal spot-market models} 
For a given investment vector, the competitive spot-market model maximizes utility net of operating costs while omitting pollution damages:
\vspace{-0.2cm}
\begin{align}
    & \mathbf{W}^M \left( \Delta \widebar{P}_{g}^G \right) = \max \sum_{t \in \mathcal{T}}\left( \sum_{d \in \mathcal{D}} b_{d,t}^{D} P_{d,t}^{D} - \sum_{g \in \mathcal{G}} b_{g,t}^{G} P_{g,t}^{G} \right) \label{A_eq:competetive_market_objective} \\
    \text{subject to   } & \nonumber \\
    & 0 \leq P_{g,t}^{G} \leq \widebar{P}_{g}^G + \Delta \widebar{P}_{g}^G \leftrightarrow \left( \ushort{\mu}_{g,t}, \widebar{\mu}_{g,t} \right), \; \forall g \in \mathcal{G}, t \in \mathcal{T}, \label{A_eq:competetive_market_generation_limits} \\
    & 0 \leq P_{d,t}^{D} \leq \widebar{P}_{d,t}^D \leftrightarrow \left( \ushort{\alpha}_{d,t}, \widebar{\alpha}_{d,t} \right), \; \forall d \in \mathcal{D}, t \in \mathcal{T}, \label{A_eq:competetive_market_consumption_limits} \\
    & \sum_{d \in \mathcal{D}} P_{d,t}^{D} = \sum_{g \in \mathcal{G}} P_{g,t}^{G} \leftrightarrow \left( \lambda_t \right), \; \forall t \in \mathcal{T}, \label{A_eq:competetive_market_power_balance} \\
    & -\widebar{F}_l \leq \sum_{g \in \mathcal{G}} H_{g,l}^{G} P_{g,t}^{G} - \sum_{d \in \mathcal{D}} H_{d,l}^{D} P_{d,t}^{D} \leq \widebar{F}_l \leftrightarrow \left( \ushort{\beta}_{l,t}, \widebar{\beta}_{l,t} \right), \; \forall l \in \mathcal{L}, t \in \mathcal{T}. \label{A_eq:competetive_market_line_limits}
\end{align}

For the spot-market comparison without investment, $\Delta\widebar{P}_{g}^{G}=0$ for all $g\in\mathcal{G}$. Let $e_{g,t}^{G}$ denote the linear marginal pollution-damage coefficient of generator $g$ in time step $t$. The socially optimal spot-market model includes this coefficient in the dispatch objective:
\vspace{-0.2cm} 
\begin{equation}
\label{A_eq:optimal_market_objective}
    \mathbf{W}^M \left( \Delta \widebar{P}_{g}^G \right) = \max \sum_{t \in \mathcal{T}}\left( \sum_{d \in \mathcal{D}} b_{d,t}^{D} P_{d,t}^{D} - \sum_{g \in \mathcal{G}} \left( b_{g,t}^{G} + e_{g,t}^{G} \right) P_{g,t}^{G} \right),
\end{equation}
subject to \eqref{A_eq:competetive_market_generation_limits}--\eqref{A_eq:competetive_market_line_limits}. Under the normalization $\mathbf{\Phi}_{g,t}(0)=0$, the Pigouvian tax in the linear model is
\vspace{-0.2cm} 
\begin{equation}
\label{A_eq:linear_tax}
    \mathbf{\Phi}_{g,t} \left( P_{g,t}^{G} \right) = e_{g,t}^{G}P_{g,t}^{G}, \; \forall g\in\mathcal{G},\; t\in\mathcal{T}.
\end{equation}
Thus, competitive clearing with the proposed tax in \eqref{A_eq:linear_tax} is equivalent to the socially optimal spot-market model in \eqref{A_eq:optimal_market_objective}.

\vspace{-0.2cm}
\subsection{Aggregate utility and nodal prices}
To calculate the aggregate utility, nodal prices, and investment subsidy, consider the utility-maximization problem in which the generation vector $\mathbf{P}_{t}^{G}$ is fixed:
\vspace{-0.2cm} 
\begin{equation}
    \label{A_eq:primal_utility_objective}
    \mathbf{U}_t^{PP} \left( P_{g,t}^G \right) = \max \sum_{d \in \mathcal{D}} b_{d,t}^{D} P_{d,t}^{D}, \; \forall t \in \mathcal{T}.
\end{equation}
This problem is subject to \eqref{A_eq:competetive_market_consumption_limits}, \eqref{A_eq:competetive_market_power_balance}, and \eqref{A_eq:competetive_market_line_limits}, with $\mathbf{P}_{t}^{G}$ treated as fixed. Its dual problem for each time step is
\vspace{-0.2cm} 
\begin{multline}
    \label{A_eq:dual_utility_objective}
    \mathbf{U}_t^{DP} \left( P_{g,t}^G \right) =
    \min 
    \sum_{d \in \mathcal{D}} \widebar{P}_{d,t}^D \widebar{\alpha}_{d,t}
    + \left( \sum_{g \in \mathcal{G}} P_{g,t}^{G} \right) \lambda_t
    - \sum_{l \in \mathcal{L}} \left( \widebar{F}_l + \sum_{g \in \mathcal{G}} H_{g,l}^{G} P_{g,t}^{G} \right) \widebar{\beta}_{l,t}
    \\ + \sum_{l \in \mathcal{L}} \left( \widebar{F}_l - \sum_{g \in \mathcal{G}} H_{g,l}^{G} P_{g,t}^{G} \right) \ushort{\beta}_{l,t}, \; \forall t \in \mathcal{T},
\end{multline}
subject to
\vspace{-0.2cm}
\begin{equation}
    \label{A_eq:dual_utility_Pd}
    \ushort{\alpha}_{d,t} + \widebar{\alpha}_{d,t} + \lambda_t - \sum_{l \in \mathcal{L}} H_{d,l}^{D} \left( \ushort{\beta}_{l,t} +  \widebar{\beta}_{l,t} \right) = b_{d,t}^{D} \leftrightarrow \left( P_{d,t}^{D} \right), \; \forall d \in \mathcal{D}, t \in \mathcal{T},
\vspace{-0.2cm}
\end{equation}
\vspace{-0.2cm}
\begin{equation}
    \label{A_eq:dual_utility_demand_sign}
    \ushort{\alpha}_{d,t} \leq 0, \widebar{\alpha}_{d,t} \geq 0, \; \forall d \in \mathcal{D}, t \in \mathcal{T},
\vspace{-0.2cm}
\end{equation}
\vspace{-0.2cm}
\begin{equation}
    \label{A_eq:dual_utility_line_sign}
    \ushort{\beta}_{l,t} \leq 0,\widebar{\beta}_{l,t} \geq 0, \; \forall l \in \mathcal{L}, t \in \mathcal{T}.
\end{equation}

When $\mathbf{P}_{t}^{G}$ is fixed, the primal and dual problems are linear programs, and strong duality holds. In the complete model, however, generation is endogenous. We therefore impose primal feasibility, dual feasibility, and complementary slackness. The cases $\gamma=1.01$ and $\gamma=1.5$ produce a distinct extreme solution with 3405~MW of new investment, indicating that these scaling values are too restrictive for the present formulation. The baseline case with $\gamma=10$ yields 1920.30~MW of strategic investment, social welfare of \$82.89 million, and a welfare loss of 4.40\% relative to the socially optimal benchmark. For $\gamma\geq2$, the welfare loss remains within a comparatively narrow range of 4.30--4.57\%, although strategic investment varies from 1852.70~MW to 2127.90~MW. Thus, the qualitative conclusion of strategic overinvestment and lower welfare is robust across the accepted scaling values, but the precise equilibrium investment solution remains sensitive to the big-$M$ calibration. Recent disjunctive-based decomposition methods for linear bilevel programs can avoid explicit big-$M$ selection \citep{mohammadi2025famous}. Adapting such a parameter-free approach to the present strategic-investment formulation could remove the need to tune $\gamma$ and is therefore an interesting direction for future computational work. The complementary-slackness conditions for the demand bounds are linearized as

\vspace{-0.2cm}
\begin{equation}
    \label{A_eq:Linear_Complementary_Slackness_Utility_Pdmin_up}
    P_{d,t}^{D} \leq \gamma \widebar{P}_{d,t}^D \left( 1-\ushort{z}_{d,t}^D \right), \; \forall d \in \mathcal{D}, t \in \mathcal{T},
\vspace{-0.2cm}
\end{equation}
\vspace{-0.2cm}
\begin{equation}
    \label{A_eq:Linear_Complementary_Slackness_Utility_Pdmin_dn}
    \ushort{\alpha}_{d,t} \geq - \gamma b_{d,t}^{D} \ushort{z}_{d,t}^D, \; \forall d \in \mathcal{D}, t \in \mathcal{T},
\vspace{-0.2cm}
\end{equation}
\vspace{-0.2cm}
\begin{equation}
    \label{A_eq:Linear_Complementary_Slackness_Utility_Pdmax_dn}
    P_{d,t}^{D} \geq \widebar{P}_{d,t}^D - \gamma \widebar{P}_{d,t}^D \left( 1-\widebar{z}_{d,t}^D \right), \; \forall d \in \mathcal{D}, t \in \mathcal{T},
\vspace{-0.2cm}
\end{equation}
\vspace{-0.2cm}
\begin{equation}
    \label{A_eq:Linear_Complementary_Slackness_Utility_Pdmax_up}
    \widebar{\alpha}_{d,t} \leq \gamma b_{d,t}^{D} \widebar{z}_{d,t}^D, \; \forall d \in \mathcal{D}, t \in \mathcal{T}.
\end{equation}
The linearized complementary slackness constraints for line limits are
\vspace{-0.2cm}
\begin{equation}
    \label{A_eq:Linear_Complementary_Slackness_Utility_Flmin_up}
    \sum_{g \in \mathcal{G}} H_{g,l}^{G} P_{g,t}^{G} - \sum_{d \in \mathcal{D}} H_{d,l}^{D} P_{d,t}^{D}
    \leq -\widebar{F}_l + \left( \gamma + 1 \right) \widebar{F}_l \left( 1-\ushort{z}_{l,t}^L \right), \; \forall l \in \mathcal{L}, t \in \mathcal{T},
\vspace{-0.2cm}
\end{equation}
\vspace{-0.2cm}
\begin{equation}
    \label{A_eq:Linear_Complementary_Slackness_Utility_Flmin_dn}
    \ushort{\beta}_{l,t} \geq - \gamma \max_{d \in \mathcal{D}} \left( b_{d,t}^{D} \right) \ushort{z}_{l,t}^L, \; \forall l \in \mathcal{L}, t \in \mathcal{T},
\vspace{-0.2cm}
\end{equation}
\vspace{-0.2cm}
\begin{equation}
    \label{A_eq:Linear_Complementary_Slackness_Utility_Flmax_dn}
    \sum_{g \in \mathcal{G}} H_{g,l}^{G} P_{g,t}^{G} - \sum_{d \in \mathcal{D}} H_{d,l}^{D} P_{d,t}^{D}
    \geq \widebar{F}_l - \left( \gamma + 1 \right) \widebar{F}_l \left( 1-\widebar{z}_{l,t}^L \right), \; \forall l \in \mathcal{L}, t \in \mathcal{T},
\vspace{-0.2cm}
\end{equation}
\vspace{-0.2cm}
\begin{equation}
    \label{A_eq:Linear_Complementary_Slackness_Utility_Flmax_up}
    \widebar{\beta}_{l,t} \leq \gamma \max_{d \in \mathcal{D}} \left( b_{d,t}^{D} \right) \widebar{z}_{l,t}^L, \; \forall l \in \mathcal{L}, t \in \mathcal{T},
\vspace{-0.2cm}
\end{equation}
\vspace{-0.2cm}
\begin{equation}
    \label{A_eq:Linear_Complementary_Slackness_Utility_Binary_demand}
    \ushort{z}_{d,t}^D, \widebar{z}_{d,t}^D \in \{ 0,1 \}, \; \forall d \in \mathcal{D}, t \in \mathcal{T},
\vspace{-0.2cm}
\end{equation}
\vspace{-0.2cm}
\begin{equation}
    \label{A_eq:Linear_Complementary_Slackness_Utility_Binary_line}
    \ushort{z}_{l,t}^L, \widebar{z}_{l,t}^L \in \{ 0,1 \}, \; \forall l \in \mathcal{L}, t \in \mathcal{T}.
\end{equation}

Adding \eqref{A_eq:dual_utility_Pd}--\eqref{A_eq:Linear_Complementary_Slackness_Utility_Binary_line} to the competitive or socially optimal market model yields a MILP that provides the multipliers $(\lambda_t,\ushort{\beta}_{l,t},\widebar{\beta}_{l,t})$ and therefore the corresponding nodal prices. The price at bus $i$ is
\vspace{-0.2cm}
\begin{equation}
    \label{A_eq:Price_bus}
    \Theta_{i,t}^{I} = \lambda_t - \sum_{l \in \mathcal{L}} H_{i,l}^{I} \left( \ushort{\beta}_{l,t} + \widebar{\beta}_{l,t} \right), \; \forall i \in \mathcal{I}, t \in \mathcal{T}.
\vspace{-0.2cm}
\end{equation}
Similarly, prices at generator and demand locations are
\vspace{-0.2cm}
\begin{equation}
    \label{A_eq:Price_gen}
    \Theta_{g,t}^{G} = \lambda_t - \sum_{l \in \mathcal{L}} H_{g,l}^{G} \left( \ushort{\beta}_{l,t} + \widebar{\beta}_{l,t} \right), \; \forall g \in \mathcal{G}, t \in \mathcal{T},
\vspace{-0.2cm}
\end{equation}
\vspace{-0.2cm}
\begin{equation}
    \label{A_eq:Price_demand}
    \Theta_{d,t}^{D} = \lambda_t - \sum_{l \in \mathcal{L}} H_{d,l}^{D} \left( \ushort{\beta}_{l,t} + \widebar{\beta}_{l,t} \right), \; \forall d \in \mathcal{D}, t \in \mathcal{T}.
\end{equation}

For the subsidy calculation, let $\mathbf{U}_{t}^{(-g)}$ denote the value of \eqref{A_eq:primal_utility_objective} when generator $g$ is fixed at zero output and the remaining market is re-cleared. Under the normalization $\mathbf{\Psi}_{g,t}(0)=0$, the linear-model subsidy is
\vspace{-0.2cm}
\begin{equation}
\label{A_eq:linear_subsidy}
    \mathbf{\Psi}_{g,t} = \mathbf{U}_{t}^{PP} \left( \mathbf{P}_{t}^{G} \right) - \mathbf{U}_{t}^{(-g)} - \Theta_{g,t}^{G}P_{g,t}^{G}, \; \forall g\in\mathcal{G},\; t\in\mathcal{T}.
\end{equation}
Thus, computing the subsidy requires one additional counterfactual market-clearing problem for each generator and time step at which the subsidy is evaluated.

\vspace{-0.2cm}
\subsection{Optimal and strategic capacity investment}
Let $c_{g}^{G}$ denote the annualized investment cost per MW \citep{eia2022cost}, and let $\Delta\widebar{P}_{g}^{G\max}$ denote the upper bound on new capacity. The socially optimal investment problem is
\vspace{-0.2cm}
\begin{equation}
    \label{A_eq:Optimal_Investment_objective}
    \mathbf{W}^I = \sum_{t \in \mathcal{T}}\left( \sum_{d \in \mathcal{D}} b_{d,t}^{D} P_{d,t}^{D} - \sum_{g \in \mathcal{G}} \left( b_{g,t}^{G} + e_{g,t}^{G} \right) P_{g,t}^{G} \right) - \sum_{g \in \mathcal{G}} c_{g}^{G} \Delta \widebar{P}_{g}^G,
\vspace{-0.2cm}
\end{equation}
subject to \eqref{A_eq:competetive_market_generation_limits}--\eqref{A_eq:competetive_market_line_limits},
\eqref{A_eq:dual_utility_Pd}--\eqref{A_eq:Linear_Complementary_Slackness_Utility_Binary_line}, and
\vspace{-0.2cm}
\begin{equation}
    \label{A_eq:Investment_capacity_limitation}
    0 \leq \Delta \widebar{P}_{g}^G \leq \Delta \widebar{P}_{g}^{G\max} \leftrightarrow ( \ushort{\tau}_{g}, \widebar{\tau}_{g} ), \; \forall g \in \mathcal{G}.
\end{equation}
The optimal-investment formulation may also include \eqref{A_eq:dual_utility_Pd}--\eqref{A_eq:Linear_Complementary_Slackness_Utility_Binary_line} when nodal-price and aggregate-utility variables are required.

In the strategic investment model, each generator chooses its own capacity investment while anticipating the market-clearing outcome. Generator $g$'s investment choice is represented by
\vspace{-0.2cm}
\begin{equation}
\label{A_eq:Profit_maximization_investment}
    \Delta \widebar{P}_{g}^G =
    \underset{{0 \leq \Delta \widebar{P}_{g}^G \leq \Delta \widebar{P}_{g}^{G\max}}}{\text{argmax}}
    \left[ \sum_{t \in \mathcal{T}}
    \left( \Theta_{g,t}^{G} - b_{g,t}^{G} - e_{g,t}^{G} \right) P_{g,t}^{G}
    - c_{g}^{G} \Delta \widebar{P}_{g}^G \right], \; \forall g \in \mathcal{G}.
\end{equation}
Here, dispatch and prices are determined endogenously by the market-clearing model for the resulting installed-capacity vector. The term $e_{g,t}^{G}P_{g,t}^{G}$ represents the linear Pigouvian tax.

To obtain a single-level formulation, the market-clearing system is combined with the KKT conditions associated with each generator's investment subproblem. The corresponding dual representation used in the numerical formulation is
\vspace{-0.2cm}
\begin{align}
    & \min \sum_{t \in \mathcal{T}} \left( \widebar{P}_{g}^G - P_{g,t}^{G} \right) \widebar{\mu}_{g,t}^G + \Delta \widebar{P}_{g}^{G\max} \widebar{\tau}_{g}, \label{A_eq:dual_profit_maximization_investment_objective} \\
    \text{subject to} \quad \quad \quad \; \; & \nonumber \\
    & \ushort{\tau}_{g} + \widebar{\tau}_{g} - \sum_{t \in \mathcal{T}} \widebar{\mu}_{g,t}^G = - c_{g}^{G} \leftrightarrow \left( \Delta \widebar{P}_{g}^G \right), \; \forall g \in \mathcal{G}, \label{A_eq:dual_profit_maximization_investment_delta_Pg} \\
    & \ushort{\tau}_{g} \leq 0, \widebar{\tau}_{g} \geq 0, \; \forall g \in \mathcal{G}, \label{A_eq:dual_profit_maximization_investment_mu_Pg} \\
    & \widebar{\mu}_{g,t}^G \geq 0, \; \forall g \in \mathcal{G}, t \in \mathcal{T}.  \label{A_eq:dual_profit_maximization_investment_mu_Pgt}
\end{align}
The complementary-slackness conditions for the investment bounds and the generation-capacity constraint are linearized as
\vspace{-0.2cm}
\begin{equation}
    \label{A_eq:Linear_Complementary_Slackness_DeltaPg_up}
    \Delta \widebar{P}_{g}^G \leq \gamma \Delta \widebar{P}_{g}^{G\max} \left( 1-\ushort{z}_{g}^{\Delta G} \right), \; \forall g \in \mathcal{G},
\vspace{-0.2cm}
\end{equation}
\vspace{-0.2cm}
\begin{equation}
    \label{A_eq:Linear_Complementary_Slackness_DeltaPg_dn}
    \Delta \widebar{P}_{g}^G \geq \Delta \widebar{P}_{g}^{G\max} - \gamma \Delta \widebar{P}_{g}^{G\max} \left( 1-\widebar{z}_{g}^{\Delta G} \right), \; \forall g \in \mathcal{G},
\vspace{-0.2cm}
\end{equation}
\vspace{-0.2cm}
\begin{equation}
    \label{A_eq:Linear_Complementary_Slackness_muPgmin_dn}
    \ushort{\tau}_{g} \geq - \gamma c_{g}^{G} \ushort{z}_{g}^{\Delta G}, \; \forall g \in \mathcal{G},
\vspace{-0.2cm}
\end{equation}
\vspace{-0.2cm}
\begin{equation}
    \label{A_eq:Linear_Complementary_Slackness_muPgmin_up}
    \widebar{\tau}_{g} \leq \gamma c_{g}^{G} \widebar{z}_{g}^{\Delta G}, \; \forall g \in \mathcal{G},
\vspace{-0.2cm}
\end{equation}
\vspace{-0.2cm}
\begin{equation}
    \label{A_eq:Linear_Complementary_Slackness_muPgtmin_dn}
    P_{g,t}^{G} - \widebar{P}_{g}^G - \Delta \widebar{P}_{g}^G
    \geq - \gamma \left( \widebar{P}_{g}^G + \Delta \widebar{P}_{g}^{G\max} \right) \left( 1-z_{g,t}^{G} \right),
    \; \forall g \in \mathcal{G}, t \in \mathcal{T},
\vspace{-0.2cm}
\end{equation}
\vspace{-0.2cm}
\begin{equation}
    \label{A_eq:Linear_Complementary_Slackness_muPgtmin_up}
    \widebar{\mu}_{g,t}^G \leq \gamma \max_{d \in \mathcal{D}} \left( b_{d,t}^{D} \right) z_{g,t}^{G},
    \; \forall g \in \mathcal{G}, t \in \mathcal{T},
\vspace{-0.2cm}
\end{equation}
\vspace{-0.2cm}
\begin{equation}
    \label{A_eq:Linear_Complementary_Slackness_investment_Binary_generation}
    \ushort{z}_{g}^{\Delta G}, \widebar{z}_{g}^{\Delta G} \in \{ 0,1 \}, \; \forall g \in \mathcal{G},
\vspace{-0.2cm}
\end{equation}
\vspace{-0.2cm}
\begin{equation}
    \label{A_eq:Linear_Complementary_Slackness_investment_Binary_generation_level}
    z_{g,t}^G \in \{ 0,1 \}, \; \forall g \in \mathcal{G}, t \in \mathcal{T}.
\end{equation}
Combining \eqref{A_eq:competetive_market_generation_limits}--\eqref{A_eq:competetive_market_line_limits}, \eqref{A_eq:dual_utility_Pd}--\eqref{A_eq:Linear_Complementary_Slackness_Utility_Binary_line}, \eqref{A_eq:Investment_capacity_limitation}, and \eqref{A_eq:dual_profit_maximization_investment_delta_Pg}--\eqref{A_eq:Linear_Complementary_Slackness_investment_Binary_generation_level} gives the single-level MILP representation used for the strategic investment simulations.

For the linear lower-level problems considered here, primal feasibility, dual feasibility, and complementary slackness are necessary and sufficient for optimality. The big-$M$ formulation is therefore an exact representation of the associated KKT system only when the bounds imposed on all primal and dual variables are valid. Exact representation of the KKT system does not imply uniqueness of the resulting equilibrium; multiple feasible equilibrium solutions may exist. For each reported solution, we check primal and dual feasibility, complementarity residuals, the primal--dual objective gap of the underlying linear problems, and the final MILP optimality gap. A solution is accepted only when these measures are below the specified numerical tolerances. This ex-post verification is required in addition to the sensitivity analysis for $\gamma$.

\begin{table}[t]
    \centering
    \caption{Sensitivity of the strategic-investment solution to the big-$M$ scaling parameter.}
    \label{tab:gamma_sensitivity}

    \scriptsize
    \setlength{\tabcolsep}{6pt}
    \renewcommand{\arraystretch}{1.15}

    \begin{tabular}{rrrrr}
        \toprule
        $\boldsymbol{\gamma}$
        & \makecell{\textbf{Runtime}\\\textbf{(s)}}
        & \makecell{\textbf{New strategic}\\\textbf{investment (MW)}}
        & \makecell{\textbf{Social}\\\textbf{welfare (\$)}}
        & \makecell{\textbf{Welfare}\\\textbf{loss (\%)}} \\
        \midrule
        1.01 & 14.07 & 3405.00 & 79,600,929.61 & 8.19 \\
        1.50 & 15.27 & 3405.00 & 79,600,929.61 & 8.19 \\
        2    & 75.89 & 2127.90 & 82,838,246.16 & 4.45 \\
        5    & 78.43 & 1885.50 & 82,841,723.92 & 4.45 \\
        8    & 28.33 & 1852.70 & 82,967,888.66 & 4.30 \\
        10   & 29.12 & 1835.30 & 83,013,882.06 & 4.25 \\
        100  & 32.29 & 1932.97 & 82,847,278.69 & 4.44 \\
        1000 & 31.30 & 2003.63 & 82,733,388.70 & 4.57 \\
        \bottomrule
    \end{tabular}

    \vspace{0.1cm}
    \begin{minipage}{0.94\linewidth}
        \footnotesize
        The welfare loss is calculated relative to socially optimal welfare as $100(W^{\mathrm{opt}}-W^{\mathrm{str}})/W^{\mathrm{opt}}$.
    \end{minipage}
\end{table}




\bibliographystyle{elsarticle-num} 
\bibliography{ref.bib}

\end{document}